\documentclass[12pt]{article}
\pdfoutput=1

\usepackage{amssymb,amsmath,mathrsfs,epstopdf,slashed,color}
\usepackage{tikz}
\usetikzlibrary{matrix}
\usetikzlibrary{positioning}
\usepackage{fancyhdr} 
\usepackage{lastpage} 
\usepackage{extramarks} 
\usepackage{graphicx} 
\usepackage{lipsum} 
\usepackage{amsmath}
\usepackage{amsfonts}
\usepackage{amssymb}
\usepackage{latexsym}
\usepackage{color}
\usepackage{xypic}
\usepackage{cancel,slashed}
\usepackage[linktocpage]{hyperref}
\usepackage{booktabs}
\usepackage{caption}
\usepackage{comment}
\usepackage{csquotes} 

\usepackage{graphicx}
\usepackage{placeins}
\usepackage{hyperref}
\usepackage{cite}

\def\calb         {{\cal B}}

\def\calo         {{\cal O}}

\def\calq         {{\cal Q}}

\def\calw         {{\cal W}}

\allowdisplaybreaks[1]

\makeatletter

\@addtoreset{equation}{section}
\makeatother

\newcommand{\vol}{\mathrm{vol}}

\title{\bf String Landscape and Fermion Masses}
\author{Stefano Andriolo, Shing Yan Li and S.-H. Henry Tye}

\begin{document}

\begin{titlepage}

\setcounter{page}{0}
  
\begin{flushright}
 \small
 \normalsize
\end{flushright}

\vskip 2.6cm
\begin{center}

{\Large \bf String Landscape and Fermion Masses}


\vskip 1.6cm
  
{\large Stefano Andriolo${}^{1,2}$, Shing Yan Li${}^2$  and S.-H. Henry Tye${}^{1,2,3}$}
 
 \vskip 0.6cm

 ${}^1$ Jockey Club Institute for Advanced Study, Hong Kong University of Science and Technology \\
 ${}^2$ Department of Physics, Hong Kong University of Science and Technology, Hong Kong\\
 ${}^3$ Laboratory for Elementary-Particle Physics, Cornell University, Ithaca, NY 14853, USA

 \vskip 0.4cm

Email: \href{mailto: sandriolo@connect.ust.hk, syliah@connect.ust.hk, iastye@ust.hk}{sandriolo at connect.ust.hk, syliah at connect.ust.hk, iastye at ust.hk}

\vskip 0.9cm
  
\abstract{\normalsize
Besides the string scale, string theory has no parameter except some quantized flux values;
and the string theory Landscape is generated by scanning over discrete values of all the flux parameters present. We propose that a typical (normalized) probability distribution $P(\calq)$ of a physical quantity $\calq$ (with nonnegative dimension) tends to peak (diverge) at $\calq=0$ as a signature of string theory. In the Racetrack K\"ahler uplift model, where $P(\Lambda)$  of the cosmological constant $\Lambda$  peaks sharply at $\Lambda=0$, the electroweak scale (not the electroweak model) naturally emerges when the median $\Lambda$ is matched to the observed value. We check the robustness of this scenario. In a bottom-up approach, we find that the observed quark and charged lepton masses are consistent with the same probabilistic philosophy, with distribution $P(m)$ that diverges at $m=0$, with the same (or almost the same) degree of divergence. This suggests that the Standard Model has an underlying string theory description, and yields relations among the fermion masses, albeit in a probabilistic approach (very different from the usual sense). Along this line of reasoning, the normal hierarchy of neutrino masses is clearly preferred over the inverted hierarchy, and the sum of the neutrino masses is predicted to be $\sum m_{\nu} \simeq 0.0592$ eV, with an upper bound  $\sum m_{\nu} <0.066$ eV. This illustrates a novel way string theory can be applied to particle physics phenomenology.

} 
   
\vspace{0.3cm}
\begin{flushleft}
 \today
\end{flushleft}
 
\end{center}
\end{titlepage}

\setcounter{page}{1}
\setcounter{footnote}{0}

\tableofcontents

\parskip=5pt

\section{Introduction}
\label{Sec:intro}

In Science, a simple criteria on the success of a model/idea is the number of parameters needed to explain/predict the observable phenomena. With about 20 parameters (the number of parameters in the neutrino sector remains open), the Standard Model (SM) of strong and electro-weak interactions has been spectacularly successful in fitting/predicting thousands of data points. Can we find a theory that can further reduce the 20 or so parameters? Or equivalently, can we find relations among the parameters? It is not unthinkable that string theory, as a consistent theory of quantum gravity with a single parameter (the string scale $M_S \simeq \alpha'^{-1/2}$), has the Standard Model (and beyond) as one of its solutions. Because of its richness, however, so far we have failed to find the SM as a solution. Therefore, we advocate that the strategy might need some changes in the approach (or viewpoint). 
In particular, we propose to apply the intrinsic probabilistic nature of string theory to guide the search for the SM inside string theory.  In a recent paper \cite{Andriolo:2018dee}, we show the emergence of a new scale (${\bf m} \sim 10^2$ GeV), which happens to be very close to the electro-weak scale. In this paper, we show how we can make predictions about neutrino masses. This novel approach in considering string theory opens a new door to do particle physics phenomenology.  

\begin{figure}[h]
\centering
\begin{minipage}{0.9\textwidth}
\centering
 \includegraphics[scale=0.7]{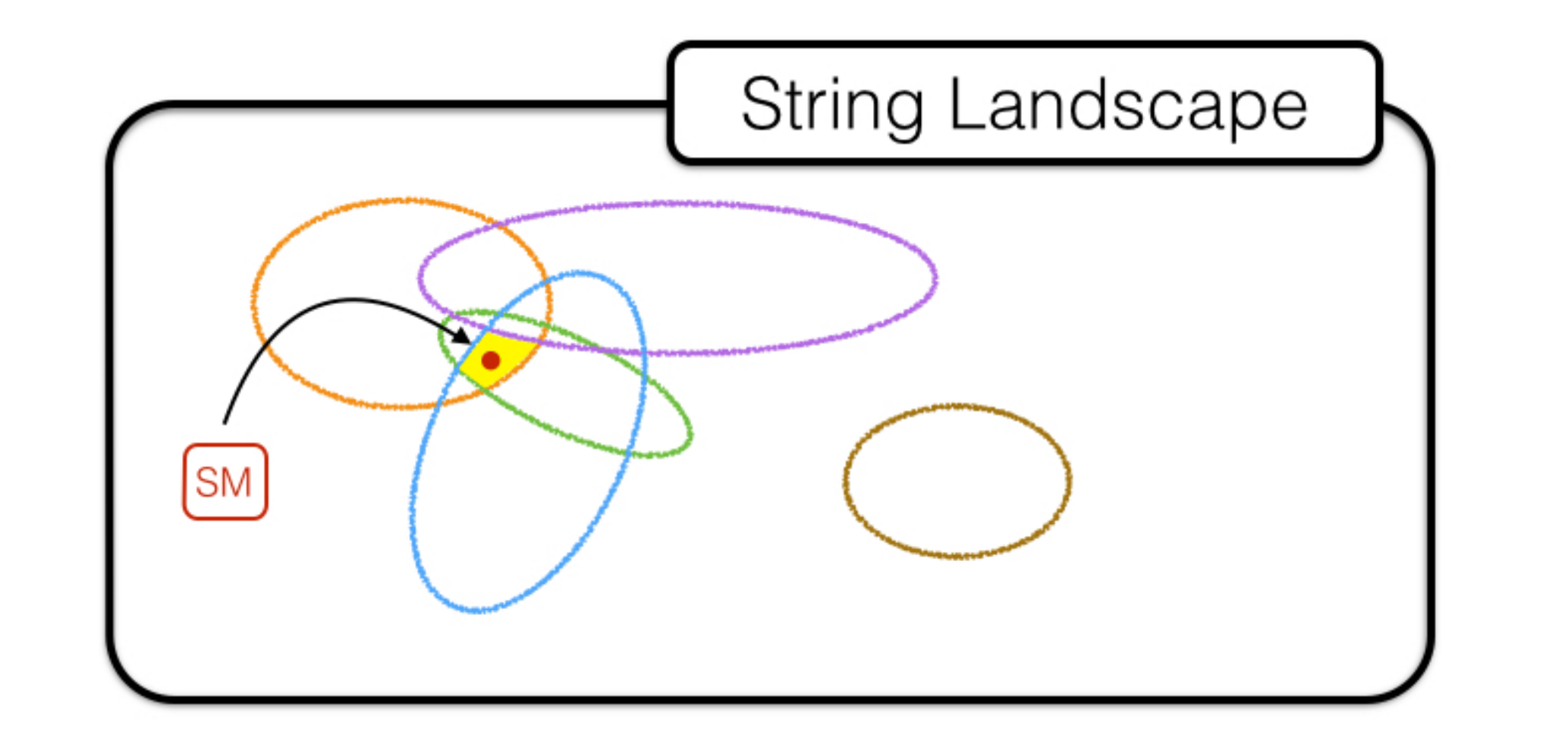}
 \end{minipage}
 \caption{\label{Fig:Landscape} \small{The Landscape of string vacua. Different string ingredients (fluxes, localised objects) give rise to ``qualitatively different'' models or patches in the string Landscape, illustrated with different colors. Different patches are described by different effective low energy theories. Moreover, given a patch, there is still some freedom, e.g., dictated by the choice of flux values. Therefore, each point corresponds to a vacuum with a specific qualitative and/or quantitative description of physics. One of these we assume is the Standard Model. The challenge is to identify as many criteria/properties as possible to guide our search of the SM in the string Landscape.}}
\end{figure}

We know that string theory allows for a Landscape. If we start with solutions that have a 4-dimensional spacetime with a dynamically flux compactified 6-dimensional manifold, scanning over all discrete flux values will yield a region of the string Landscape with an exponentially large amount of possibilities. At the low energy 4-dim effective field theory level, this corresponds to the introduction of a potential $V(F^i, \phi_j)$, where the flux parameters $F^i$ take discrete values and the moduli (scalar fields) $\phi_j$ are to be stabilized dynamically at some local minima. Parts of the string Landscape are generated as we scan over the ``dense discretuum'' of all flux values. Note that the number of flux parameters and fields may differ in different regions of the Landscape. As we move in the field space, a heavy mode that has been integrated out can become light, so we can no longer integrate it out, and we must include it into the effective potential. Additional $F^i$ must be introduced when new cycles appear. In short, a given $V(F^i, \phi_j)$ corresponds to a ``model'', valid only over a patch of the Landscape, as illustrated in Fig.~\ref{Fig:Landscape}.
Our fundamental assumption here is that string theory contains a solution that describes our universe today.

In a patch of the Landscape, before fixing $M_P$ or $\Lambda$, string theory has no scale except the string scale $M_S$, so the probability distribution of any discrete flux parameter should be flat (see discussion below on this point).
 Our proposal is that many physical quantities are determined probabilistically in the Landscape. In the absence of parameters, the probability distribution $P(\calq)$ for a quantity $\calq(F^i)$ ($\ll 1$ if it has no dimension, or well below the string scale if it has a positive dimension), is either flat or peaking (diverging) at $\calq=0$; otherwise, in the absence of fine-tuning, a scale other than the string scale will be introduced. Based on observations, it is then believed that the peaking or divergence at $\calq\to0$ is quite generic. The degree of divergence at $\calq\to0$ is determined by dynamics, and should be independent of the scale which will be introduced when we have more information, as illustrated with the relation between $\Lambda$, $M_P$ and $\bf m$ \cite{Andriolo:2018dee}. 
 The divergent behavior of $P(\calq)$ is natural in the context of string theory and is taken as a signature of string theory. Applying this to the cosmological constant $\Lambda$ allows us to understand why an exponentially small $\Lambda$ can be natural \cite{Sumitomo:2013vla}. In this paper, we show that the distribution of fermion masses reveals a distribution $P(m)$ that peaks (diverges) at $m=0$, which we interpret as evidence that the SM has an underlying string theory description. Applying this observation to the neutrinos allows us to 
see that the normal hierarchy is strongly preferred over the inverted hierarchy, and obtain a prediction on the sum of neutrino masses,
\begin{align}
\label{neutrinos}
m_1 \simeq 10^{-8} \, \text{eV}, \quad \quad \sum_{\nu=1,2,3} m_\nu \simeq 0.0592^{+0.0005}_{-0.0004} \, \text{eV},
\end{align}

Moreover, following the peaking behavior of $P(m)$, we can obtain (numerically) the approximate mass dependence on the fluxes $m(F_i)$. 
So in the search for the SM in the Landscape, we look for a specific model with the right gauge group and particle content, in which $m(F_i)$ is reproduced. This illustrates how the bottom-up approach may reveal properties of the environment where the SM sits in the Landscape. Combining the top-down and the bottom-up approaches hopefully offers new strategies in finding where the SM is hiding in the Landscape, the holy grail of fundamental physics.


We have already advocated a probability distribution approach to address the cosmological constant issues (see \cite{Tye:2016jzi} and references therein). Starting with a low energy effective theory derived (top-down) and/or inspired by string theory, we can determine $\Lambda$ in terms of $M_P$, since both are calculable from $M_S$. By scanning over the discrete flux values, one can obtain the probability distribution $P(\Lambda)$ for $\Lambda$. If the median value is comparable to the observed value $\Lambda_{obs}$, then we consider the smallness of $\Lambda_{obs}$ to be statistically natural. This happens if the properly normalized $P(\Lambda)$ peaks sharply (i.e., diverges) at $\Lambda=0$. Requiring a statistically naturally small $\Lambda$ should  be a powerful clue in the search of the SM in the Landscape.  

In particular, we observe this feature in the the K\"ahler uplift model \cite{Balasubramanian:2004uy,Westphal:2006tn,Rummel:2011cd,deAlwis:2011dp,Sumitomo:2012vx}, which is up-to-date the best controlled way to reach deSitter (dS) space among the proposed ones. In the Racetrack K\"ahler uplift model \cite{Sumitomo:2013vla}, one calculates the probability distribution $P(\Lambda)$ and finds that the median value can easily match the observed value. Furthermore, a new scale ${\bf m} \simeq10^2$ GeV automatically emerges\cite{Andriolo:2018dee}.  
It is important to explore other patches in the Landscape to check the robustness (or uniqueness) of this intriguing relation.

Similar to the statistical determination of $\Lambda$, we assume that the quark masses are not precisely calculable, but are flux-dependent $m(F_i)$ so they are described by a probability distribution $P(m)$ that peaks (diverges) at $m=0$. 
The quark mass distribution we find fits well also the charged lepton masses, when we keep the same divergent behavior  but re-set the overall energy scale, which is expected to be lower due to the absence of the strong interaction among the leptons. We like to believe that this is an improvement, since 9 fermion masses appear natural using 3 parameters for the two probability distributions. We consider this peaked probability distribution as evidence that the SM has an underlying string theory description.

The distribution of fermion masses have also been studied before \cite{Long:2017dru}. In particular, also inspired by the string theory landscape, Ref\cite{Donoghue:1997rn,Donoghue:2005cf} already pointed out that, among other properties, the measured fermion masses tend to follow distributions that peak at zero value. In this paper, motivated by the peaking behavior of $P(\Lambda)$, we focus our attention on giving a more explicit string theoretical motivation of such a peaking behavior of probability distributions. Furthermore, with more recent data, we pushed this philosophy further by providing a prediction on the value of the lightest neutrino (and so the sum of neutrino masses).

The paper is organised as follows. In Section \ref{Sec:toymodels}, we state the string theory constraints that lead to our proposal that the probability distributions of some physical quantities with positive dimensions should tend to peak (diverge) at zero value. (More discussions on this proposal can be found in the last section.) As examples, we also review and extend some basic observations given in \cite{Tye:2016jzi} regarding the peaking behaviour of $P(\Lambda)$ for some toy models.  In Section \ref{Sec:fermions}, we apply our proposal to the measured quark and charged lepton masses, and we find that a probability distribution of $P(m_f\to0) \sim m_f^{-0.731}$ gives excellent fits. We then apply this result to neutrino masses to obtain \eqref{neutrinos} for the normal hierarchy case. The Seesaw case is also considered, with almost no change in the prediction. In Section \ref{Sec:Kuplift}, we review the Racetrack K\"ahler uplift of \cite{Andriolo:2018dee} and furnish new details confirming its robustness. We show that extending the Racetrack from two to three non-perturbative terms changes only the quantitative properties but not the qualitative features. We discuss the ``attractive basin'' (field range) and comment on the ``flux basin'' (flux range) of the Racetrack K\"ahler uplift model.  We also argue that this racetrack K\"ahler uplift solution is consistent with the Swampland``distance/no de Sitter" conjecture. Section \ref{Sec:discussion} contains some discussion and conclusion. Here we also explain our proposal of the peaking $P(\calq)$ in more details. So far, the emergence of the ${\bf m}\sim 10^2$ GeV scale and the fermion mass spectra remain to be tied together. We hope to relate them in future work.

\section{Overview}
\label{Sec:toymodels}

In this Section, we describe our main proposal on the probability distributions. As illustrations, we recall some results obtained in a toy model \cite{Sumitomo:2013vla,Tye:2016jzi} and extend them to illustrate the importance and validity of our main input/assumptions. The main point is to show that if we write down an effective potential satisfying some stringy requirements, then vacua with small vacuum energy are statistically preferred. Inspired by string theory, we rely upon three well-established stringy conditions as 
\begin{description}
\item[(1)] absence of free parameters;
\item[(2)] no decoupled sector;
\item[(3a)] a discretuum of flux values with smooth distributions;\footnote{Notice that throughout the paper we will use the terms ``fluxes'' ($F_i$) and ``flux parameters/data''($a,b,c...$) interchangeably. More generally, we assume that the probability distribution of discrete values $F_i$ of fluxes is flat or smooth, while $a,b,c...$ are actually functions of $F_i$ so their probability distributions may not be smooth. In that case, we expect some of them to peak at zero values.} 
\end{description}
plus, we assume also that 
\begin{description}
\item[(3b)] the discretuum of flux data is ``dense enough'' to make a statistical analysis meaningful. Let us stress here that, without explicit calculations, we do not know ``how dense'' the discretuum is.
\end{description}
Then, as stated in Section \ref{Sec:intro} and explained more below, the probability distribution $P(\calq)$ for a low-energy physical quantity $\calq$ with nonnegative dimension peaks at $\calq=0$.

For instance, a simple toy model 
obeying (1)--(3) is \cite{Tye:2016jzi}:
\begin{equation}
\label{toymodel}
V(\phi) = \sum^n_{k=1}\frac{ a_k}{k !}\phi^k \,,
\end{equation}
with real scalar field $\phi$ and independent (real) flux parameters $a_k$'s. Given $V(\phi)$, we can compute the vacuum energy $\Lambda$ of locally stable minima as a function of flux data, $\Lambda(a_k)$, and obtain its distribution $P(\Lambda)$ after scanning over all $a_k$ values with flat or smooth distributions $P_k(a_k)$. (By a smooth probability distribution, we mean a distribution that is nowhere divergent and is non-vanishing at zero.) For simplicity, we deal with three cases, truncating the potential to $\phi^3,\phi^4,\phi^6$. The $\phi^3$ toy model can be studied analytically, while $\phi^4,\phi^6$ are studied numerically. Remarkably, one finds the same result: $P(\Lambda)$ peaks (i.e., diverges) at $\Lambda\to0^\pm$, and there is no observable difference among $\phi^3,\phi^4$ and $\phi^6$ models. (In particular, for $\phi^3$, one can analytically see that $P(\Lambda\to0^+)\sim-\log(\Lambda)$.) Moreover, this peaking behaviour is independent on the $P_k(a_k)$ chosen, as long as they are smooth \cite{Tye:2016jzi}.

As explained in \cite{Tye:2016jzi}, including two-loops radiative corrections in the $\phi^4$ case does not change $P(\Lambda)$. However, if we add an independent $a_0$ (e.g., Bousso--Polchinski terms \cite{Bousso:2000xa}) or a decoupled sector (e.g., a potential for another field not coupled to $V$) to $V(\phi)$ \eqref{toymodel}, the peaking in $P(\Lambda)$ typically disappears.
This is why condition (2) is crucial to rule out these possibilities. Fortunately, there is no decoupled sector in string theory, since all fields (moduli) couple to the closed string sector modes including the graviton and the dilaton. Furthermore, they can couple via flux parameters as well. In other words, all fields are all coupled together, directly or indirectly, via moduli and/or flux parameters.


In the Racetrack K\"ahler uplift model in Type IIB string theory, we find that $P(\Lambda)$ peaks sharply at $\Lambda=0$ \cite{Sumitomo:2013vla}. Note that both the K\"ahler uplift model \cite{Balasubramanian:2004uy,Westphal:2006tn,Rummel:2011cd,deAlwis:2011dp,Sumitomo:2012vx} and the Racetrack model \cite{Krasnikov:1987jj,Taylor:1990wr,Denef:2004dm} are scenarios well explored in string phenomenology. Matching the median $\Lambda_{50}$ to the observed value, we find that the electro-weak scale (${\bf m} \simeq 10^{2}$ GeV) emerges automatically \cite{Andriolo:2018dee}, without knowing anything about the electro-weak model. This result rests on a F-term effective potential $V$ in the low energy supergravity framework. For electroweak phenomenology, one may like to introduce a D-term (or some other term) to the potential. In the field theory framework, such a term will shift the vacuum energy density by a value orders of magnitude bigger than the observed $\Lambda$, thus ruining the above small $\Lambda$ property. However, in string theory, such a new term must couple to the moduli and/or flux parameters in the F-term $V$. Coupling the two terms together and going to the resulting minimum typically renders them to have comparable magnitudes. So it is plausible that the introduction of new terms to $V$ will not ruin the peaking behavior of $P(\Lambda)$, only shift $\Lambda$ and ${\bf m}$ by no more than a few orders of magnitude, thus not spoiling the remarkable result relating $\Lambda$ and ${\bf m}$.  It is important to find out how a D-term (or some other term) for the electroweak interaction couples to the moduli/flux parameters in the F-term $V$. This should put a tight constraint on the possible origin of such terms.

In principle, we could have applied the same statistical approach to other uplift models, such as the KKLT mechanism \cite{Kachru:2003aw}. Unfortunately, not enough is understood about the properties of the KKLT model to allow us to carry out a meaningfully reliable analysis. It is important to study those possibilities more carefully, in particular how the uplift term couples to the closed string modes and the fluxes. Recent results \cite{Gautason:2019jwq,Hamada:2018qef,Carta:2019rhx} may help in this direction.

Notice that a crucial working assumption was (3b), that is taking the flux discretuum to be ``dense enough" in order to deal with smooth (i.e., quasi-continuous) flux distributions.
At this point, one should wonder if string theory really allows for smooth flux distributions. Or, in other words, is (3b) a good assumption?
Indeed, if the discretuum is not dense enough, dS vacua might be disallowed. Let us illustrate this fact with the simple case of $\phi^3$ model, which can be dealt with analytically. There, we can set a scale so that $a_3=1$. By imposing $V'=0,V''>0$ we find the minimum and its vacuum energy:
\begin{align}
\phi_{min}=-a_2+\sqrt{a_2^2-2a_1} \,, \quad \Lambda=-\frac{(a_2 - \sqrt{a_2^2-2a_1})^2 (a_2+2\sqrt{a_2^2-2a_1})}{6} \,,
\end{align}
with the condition $a_2^2-2a_1>0$. Therefore, for a $\Lambda >0$ solution to exist, we require
\begin{align}
\label{fluxbasin}
a_2 <0 \,, \qquad \frac{3}{8}a_2^2 < a_1 < \frac{a_2^2}{2} \,.
\end{align}
We will refer to this region as the ``flux basin''. If, instead of fixing $a_3=1$, we allow $a_3$ to take discrete values, the above analysis is still valid if we replace $a_1 \to a_1/a_3$  and $a_2 \to a_2/a_3$ in the above equations. Note that the scale changes, but the degree of divergence  $P(\Lambda\to0^+)\sim-\log(\Lambda)$ remains intact.

More generally speaking, a flux basin is the range of discrete flux parameter choices $a_k$ within which the minimum behaves qualitatively the same. To be more specific, we have in mind that in the neighbourhood of  the SM in the Landscape, the particle content is the same but the quantitative values of the 20 or so parameters may vary as we vary the flux values within the flux basin. We shall provide evidence that such a flux basin exists in the vicinity of the SM in the Landscape.

\begin{figure}[h!]
\centering
\begin{minipage}{0.5\textwidth}
\centering
 \includegraphics[scale=0.45]{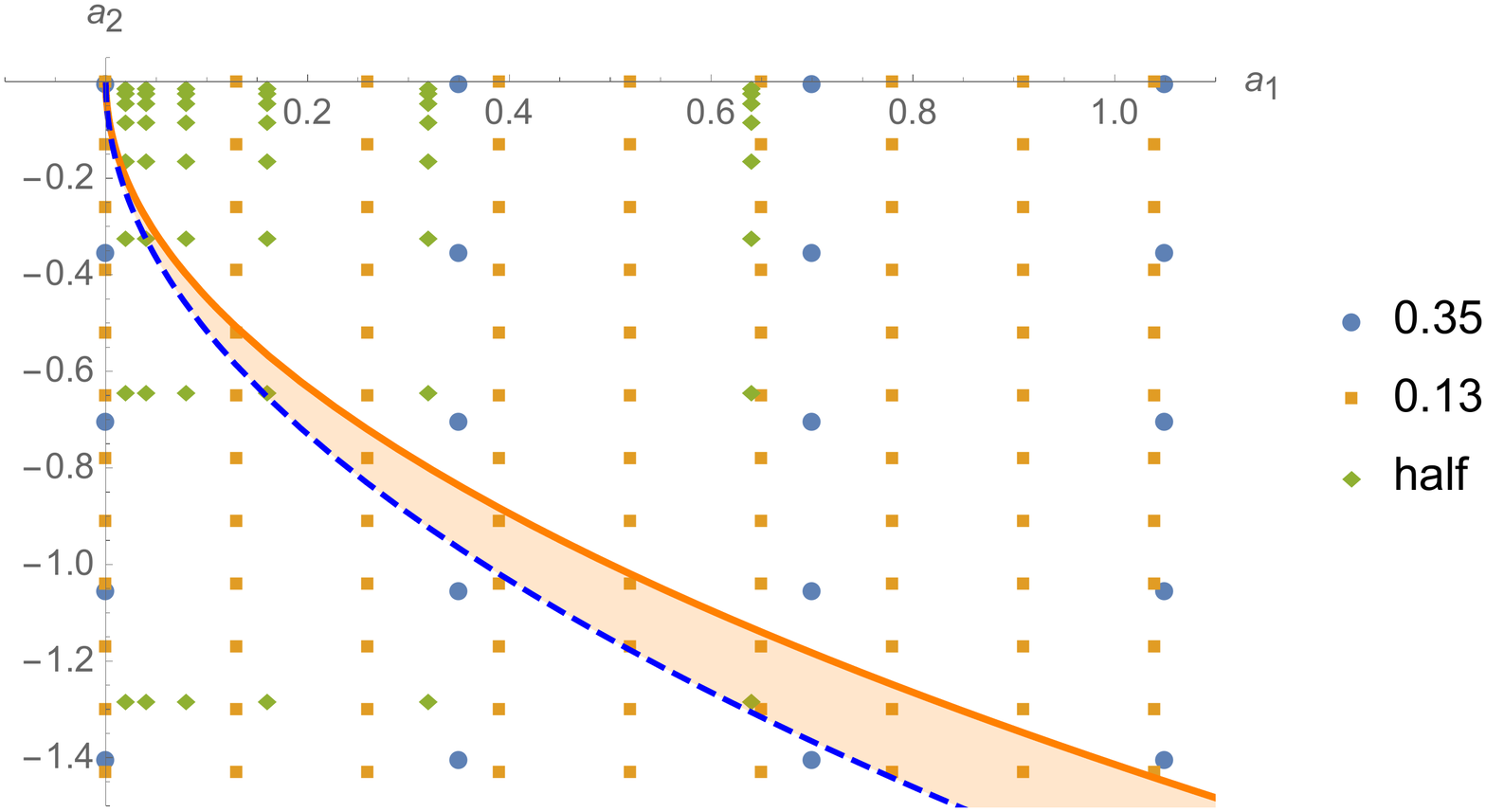}
 \end{minipage}
 \caption{\label{Fig:fluxbasin3} \small{Flux parameters $(a_1,a_2)$ take discrete values. Only $(a_1,a_2)$ belonging to the flux basin \eqref{fluxbasin} (the orange shaded area) give dS minima. The $(a_1,a_2)$ discretuum must be dense enough to allow for dS vacua. For instance, blue and orange points are obtain with uniform distributions with spacings $0.35,0.13$ respectively, while green points are obtained with a non-uniform distribution where spacings are doubled starting from a minimum spacing of $0.02$ for $a_1$ and $0.01$ for $a_2$. It is clear that the orange one allows for more dS vacua than the other two. They must be denser for $P(\Lambda)$ to be statistically meaningful.}}
\end{figure}

In the $\phi^3$ model, the better the distributions $P(a_1),P(a_2)$ fill the flux basin, the more dS vacua we obtained. There are several possibilities, as illustrated with a few examples in Fig.~\ref{Fig:fluxbasin3}. It may be possible that the separations between discrete flux values is so big that no dS solution exists (green and blue points). It is also possible that the discretuum is barely dense enough so the flux basin is not empty (orange points).
We assume that the discretuum is dense enough that a smooth probability distribution is meaningful. For instance, Ref.~\cite{Bousso:2000xa} argues that 14  flux parameters should be enough to provide a dense enough discretuum to accommodate the observed $\Lambda$. This is not difficult to achieve in flux compactifications in string theory, though one cannot be sure in the absence of an explicit construction.
Having a dense discretuum (3b) is crucial in the more realistic model of Racetrack K\"ahler uplift \cite{Andriolo:2018dee} in order to explain how the median value $\Lambda_{50}$ can be as small as the observed value. Moreover, as we will see in Section \ref{Sec:fermions}, (3b) allows us to draw important conclusions also regarding the probability distribution of fermion masses, which in turn supports the dense discretuum picture a posteriori. 

\begin{figure}[h!]
\centering
\begin{minipage}{0.5\textwidth}
\centering
 \includegraphics[scale=0.3]{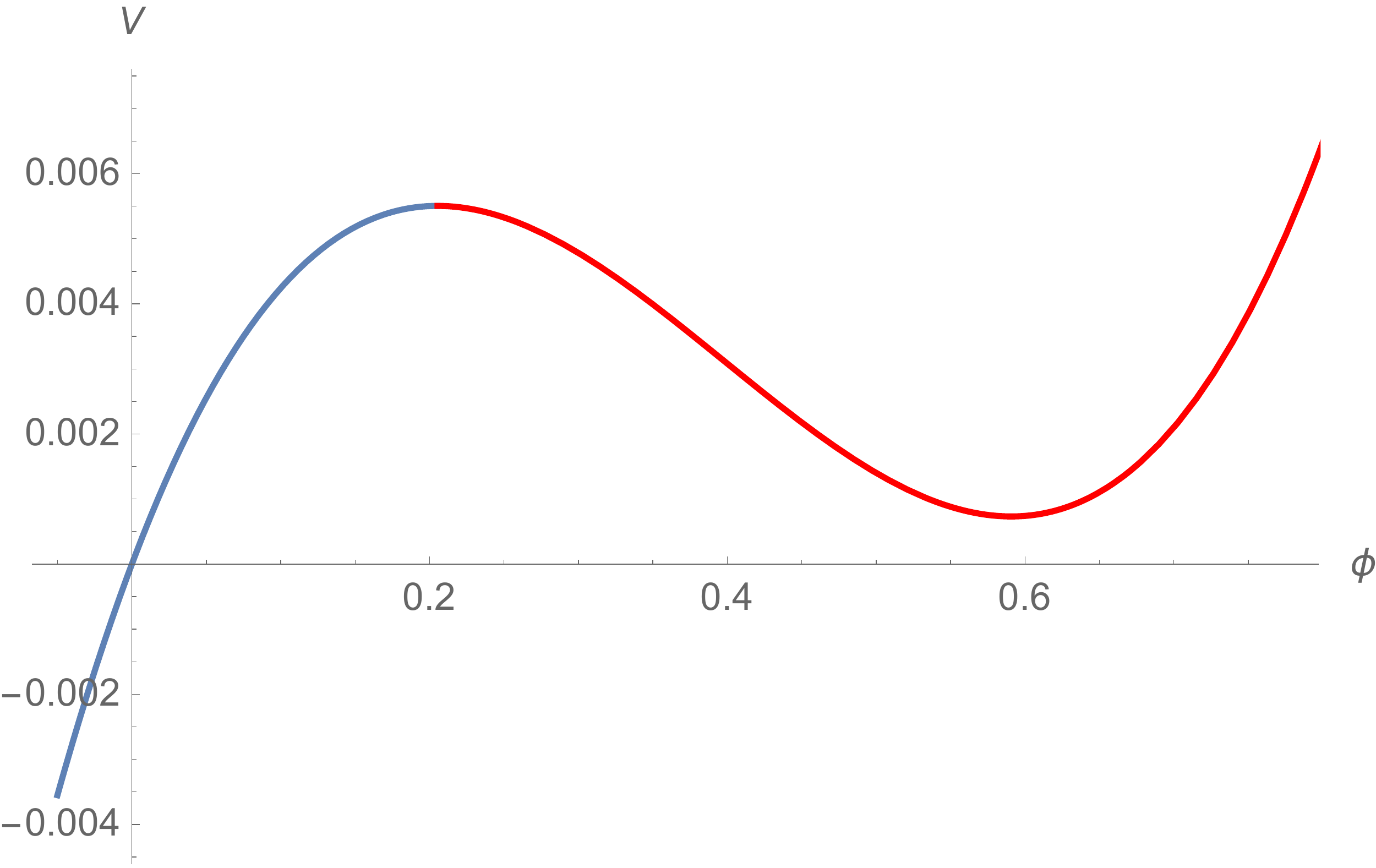}
 \end{minipage}
 \caption{\label{Fig:attractivebasin3} \small{The attractive basin $\calb=[\phi_{max},\infty]$ is the red region.}}
\end{figure}

Before explaining our approach, let us introduce here another notion, which we will use in the following. We define the ``attractive basin'', $\calb$, as the region in the field space around $\phi_{i, min}$ such that if $\phi_i  \in\calb$, then each $\phi_i$ will roll towards $\phi_{i, min}$. For instance, in the case of $\phi^3$ (with $a_3=1$), the attractive basin is simply (see Fig.~\ref{Fig:attractivebasin3})
\begin{align}
\label{attractivebasin}
\calb=[\phi_{max},\infty] \,, \quad \phi_{max}=-a_2-\sqrt{a_2^2-2a_1} \,.
\end{align}
That is, starting from any point inside the attractive basin, $\phi$ will roll towards the bottom. (The Hubble parameter in the expansion of the universe may help to dampen any over-shooting in the rolling. )

\section{Fermion Masses}
\label{Sec:fermions}

Based on the picture we have now, locating the Landscape region where the SM lies is not enough to allow us to calculate any of its 20 or so parameters. The existence of a SM flux basin implies that slight changes of some of the flux values most likely will yield the same model but with different values for the SM parameters. This seems to mean that the parameters of the SM may never be unambiguously determined, so finding the SM solution in string theory does not really improve our understanding of the fundamental stringy features of particle physics. 
So a new criteria or strategy is necessary in our search. As illustrated by the case of $\Lambda$ in our approach, even if it will never be precisely determined, one can find its probability distribution and check whether the observed value is natural or not. Here we like explore what we can learn via a bottom-up statistical approach. 

If probability distributions in flux compactification dictate some part of the underlying physics, we should find other evidence of this behavior. Hopefully, this will provide new hints in where to find the Standard Model in the Landscape. Here we examine the fermion masses. To be more specific, consider the small Yukawa couplings $Y_\alpha$, where $\alpha$ runs over fermions. Each $Y_\alpha$ is a function of common fluxes $F_j$ and fluxes $F_i^{(\alpha)}$ that are associated specifically to the $\alpha$-th fermions, therefore
$$Y_\alpha(F_j, F_i^{(\alpha)}) \to  Y_\alpha(F_i^{(\alpha)})$$
since $F_j$ take same values for each fermion.\footnote{In principle, $Y_\alpha$'s can be functions of some moduli $\Phi_a$ as well. However, moduli vacuum expectation values (vevs) are determined by flux data, so we can just think of the Yukawa couplings in terms of common and specific flux dependence, $Y_\alpha(F_j, F_i^{(\alpha)})$.} 
There are 3 obvious possibilities : \\
$\bullet$ In the absence of $\alpha$th-dependent $F_i^{(\alpha)}$, all fermion masses will be identical (up to the running of the Yukawa couplings); this is certainly not the case. \\
$\bullet$  The $Y_\alpha(F_i^{(\alpha)})$ has a different functional form for each $\alpha$. In which case, the actual fermion mass spectra shed little light to the underlying physics. \\
$\bullet$ It is reasonable to assume that $Y_\alpha(F_i^{(\alpha)})$ has the same functional form for each $\alpha$. Hence, if all fluxes have the same distribution (as we shall assume throughout the paper), $Y_\alpha$'s have the same $P(Y_\alpha)$. This immediately follows from the fact that
\begin{align}
P_\alpha(Y_\alpha)= \int \Pi_{i,j} dF_j dF_i^{(\alpha)} P_j(F_j) P_i(F_i^{(\alpha)})\  \delta [Y_\alpha-Y_\alpha(F_j, F_i^{(\alpha)})]  = P(Y_\alpha)\,.
\end{align}
If this is the case, any number of Yukawa couplings we sample from $P(Y_\alpha)$ should distribute accordingly. That is, the fermions obeys the same probability distribution.

In fact, we find that the distributions of both the quark and charged lepton masses closely obey a properly normalized probability distribution $P(m_q)$ that diverges at $m=0$, that is, $P(m \to 0) \propto m^{-0.731}$ , even though the quark masses and the charged lepton masses have different scales. 
This is in agreement with our probabilistic proposal. In turn, this fermion mass property sheds light on how the standard model is realized within string theory. We like to believe that this is an improvement, since 9 fermion masses appear natural using 3 parameters, namely the power of divergence, the quark mass scale $M_Q$ and the lepton mass scale $M_L$.

Here we show that both quark masses and charged lepton masses appear to be natural under probability distributions diverging at $m=0$. To be precise, under such distributions the masses have quasi-uniformly distributed percentiles as shown below. If we randomly sample the same number of fermion masses from the distributions, the chance of getting similar mass hierarchies is considerably high. To quantify such distributions, we approximate them by some well-known distributions. In particular, we choose the (properly normalized) 
Weibull distribution,
\begin{equation}
 \label{Weibullu}
 f(u;k)=ku^{k-1} e^{-u^k}, \quad \quad  \int_0^{\infty} f(u;k) \, du =1 
\end{equation}
with the shape parameter, as shown below, $k\sim0.269$ for both quark and charged leptons. For $0 <  k<1$, as is the case here, $f(u;k)$ \eqref{Weibullu} diverges at $u=0$, and $k$ measures the degree of divergence there. Here,  $f(u;k)$ depends only on the degree of divergence $0< k <1$ and has no scale. It is a prime example of what we have in mind for a typical probability distribution $P(\calq)$, which diverges at zero, is monotonic and has no scale. A scale appears only when we understand the dynamics and $\calq$ has a dimension. However, the Weibull distribution may be replaced by any distribution satisfying the above properties. See below for more discussions on the choice of the Weibull distribution.

Note that the Weibull distribution can be re-written as a function of fermion mass $x$, where $u=x/l$ and $k$ only,
\begin{equation}
 \label{Weibull}
f(x;k,l)=\frac{k}{l}\left(\frac{x}{l}\right)^{k-1} \exp \left(-\left(\frac{x}{l}\right)^k \right), \quad \quad  \int f(x;k,l) \, dx =1 
\end{equation}
 for (different) mass scale $l$, where $l$ measures the width of the divergent peak.
 For fermion masses $m$ below, the scale $l$ is obtained from fitting the data, where quarks and leptons have different $l$.
Once dynamics introduces a new scale (e.g., $\bf m$), it will in principle fix $l$, while $k$ is unchanged.
So it enjoys the feature that the median $x_{50}=l (\ln 2)^{1/k}$ is much smaller than the mean $\bar x=l \Gamma (1+ 1/k)$, 
\begin{equation}
\label{peaking}
r \equiv \frac{x_{50}}{\bar x} = \frac{(\ln 2)^{1/k}}{\Gamma (1+ 1/k)} \ll 1 \,.
\end{equation}
The actual probability distribution $P(\calq)$ for a physical quantity $\calq$ probably varies from $f(\calq/l ; k)$. For example, the divergence may involve powers of logarithms which are subdominant (e.g., see $P(\Lambda)$ \cite{Sumitomo:2013vla}), 
and $P(\calq)$  may tail off for large $\calq$ differently from $f(\calq/l ; k)$.  As we have seen \cite{Sumitomo:2013vla},  the median is more informative about string theory properties than the average value, and the former is relatively insensitive to the tail (large $u$) behavior. Numerically, it turns out that $f(x;k,l)$ \eqref{Weibull} provides an excellent fit to the fermion spectra, and varying $f(\calq/l ; k)$ without introducing a new scale does not improve the fit. In any case, varying $f(x;k,l)$ without degrading the fit to data makes no difference to the picture we have here below.

We can also apply our probabilistic philosophy to the mixing angles. The CKM mixing angles for the quarks are compatible with a peaked distribution, but, if it is present, the degree of divergence is weaker, so it cannot be taken as evidence of another signature of string theory. Furthermore, the unitarity constraint on the mixing matrix (both the CKM matrix and the neutrino mixing matrix) renders the application of the probabilistic approach less obvious than the fermion mass cases. 

\subsection{Quark Masses}

The 6 current quark masses are known \cite{Tanabashi:2018oca}, and we see 
that the Weibull distribution (\ref{Weibull}) with $k=0.269$ and $l=M_Q=2290$ MeV provides an excellent fit,
\begin{equation}
 \label{PofQ}
P(m_q)\simeq f(m_q;0.269,2290\,{\rm MeV}), \quad (\chi^2=0.67, p\text{-value}=0.96) \,,
\end{equation}
with $r=0.016$.
We see that $r$ does not match the observed ratio $m_{q_{50}}/\bar m_q\simeq 0.023$, and this is because averaging the strange and charm quark masses to obtain the median is not reliable for such a steep distribution. On the other hand, allowing a range of values for the median, we have
 $$0.0032 < \frac{m_{q_{50}}}{\bar m_q} <  0.043 \,, $$ 
which is in agreement with $r=0.016$ within the limited amount of data.
We also show in Table \ref{Tab:quarks} the percentiles corresponding to the observed values. 
\begin{table}[h!]
\begin{center}
$\begin{array}{cccccc}
\toprule 
\multicolumn{3}{c}{\text{Quarks}} & \multicolumn{3}{c}{\text{Leptons}} \\
\text{Particle} & \text{Mass (MeV)} & Y\% & 
\text{Particle} & \text{Mass (MeV)} & Y\% \\
\midrule
u& 2.3 & 14.5 & e & 0.511 & 19.1 \\
d & 4.8 & 17.4 & \mu & 106 & 58.9  \\
s& 95 & 34.7 & \tau & 1777 & 85.0 \\
c & 1275 & 57.5 & {} & {} & {} \\
b & 4180 & 69.1 & {} & {} & {}  \\
t & 173210 & 95.9 & {} & {} & {} \\   
\bottomrule
\end{array}$
\end{center}
\caption{\label{Tab:quarks} Quark and lepton masses with their percentiles $Y\%$ (i.e., there is $Y\%$ probability that $m< m_{Y}$).}
\end{table}

Notice that another distribution maintaining the (qualitative) divergent behavior at $m_f=0$ but with a different tail can also provide a reasonable fit.  Any realistic model, i.e., containing the SM as a meta-stable vacuum, should be such that the functional form $m_f(a_k)$ gives a mass distribution $P(m_f)$ that peaks at zero, such that $m_{f 50}\ll \bar m_f$. For instance, as a function of a single flux parameter with a smooth distribution $P(a)$, as $m_f \to 0$, we can choose
\begin{align}
\label{powerk}
m_f(a) \simeq  a^{1/k} = a^{3.72} \,.
\end{align}
If $m_f$ is a function of more than one flux parameter, the dependence is no longer determined.  For example, one of the following two possibilities is acceptable for $m_f(a_1, a_2)$ to produce the divergent behavior in $P(m_f)$ as $a_j \to 0$ \cite{Sumitomo:2012wa},
$$m_f(a_1, a_2) \sim a_1^{1/k}a_2^s, \quad \quad s < 1/k$$ or
 $$m_f(a_1, a_2) \sim a_1^{s_1} + a_2^{s_2}, \quad \quad \frac{1}{s_1} + \frac{1}{s_2}=\frac{1}{k} . $$
In the latter case, the divergent behavior of  individual distribution $P_i(a_i)$ is weakened in the  distribution $P(m_f)$ for the sum, but not erased. String theory constraint (2) clearly prefers the first possibility.
Overall, this analysis indicates that the flux basin is big enough, i.e., the flux parameters $a_k$ for the quark masses are dense enough to cover the range of quark masses observed in nature. 
That 6 quark masses can be described by a single probability distribution (\ref{PofQ}) that involves only 2 parameters should be considered as an improvement.

\subsection{Charged Lepton Masses}

We expect that the peaking also happens in the case of the charged lepton masses $m_l$, even though we have only 3 data points. We find that the three lepton masses are also well fitted by the Weibull distribution. (Although it is not the optimum value, choosing the same $k=0.269$ gives an excellent fit.)
\begin{equation}
\label{PofL}
P(m_l)\simeq f(m_l;0.269,164\,{\rm MeV}), \quad (\chi^2=1.00, \quad p\text{-value}=0.80) \,,
\end{equation}
which has the same $r=0.016$ as in the case of quark masses, but with a smaller mass scale $l=M_L= 164$ MeV. The different value for $l$ is not unexpected, as the QCD coupling to quarks but not leptons tend to raise the quark mass scale. The validity of these fittings can be checked from the percentiles in Table \ref{Tab:quarks}. Note that $P(m_l)$ (\ref{PofL}) is a very good fit even though the observed ratio \eqref{peaking}
\begin{equation}
\label{L50}
\frac{m_{l_{50}}}{\bar m_l} =106\,{\rm MeV}/628\,{\rm MeV}=0.169 <1\,,
\end{equation}
is not close to $r=0.016$, due to the scarcity of data.

  A few comment are in order here. 
  
 $\bullet$ Although both the quark masses and the charged lepton masses obey (almost) identical probability distributions $f(u, 0.269)$ (\ref{Weibullu}), they have different mass scales, namely  $M_Q=2290$ MeV (\ref{PofQ}) and $M_L=164$ MeV (\ref{PofL}) respectively. Recall grand unified theories. Because of the QCD contribution to the running of the quark Yukawa couplings, we do expect $M_Q>M_L$. However, simple grand unified theories typically do not yield such a big difference. So this point deserves further examination.
 
 $\bullet$ One should be careful with the meaning of ``fitting data''. Below we obtain estimates of $k$ and $l$ from the fermion masses. Obviously the masses can arise from distributions with different $k$ and $l$, or even of other types. However, we say that a distribution fits the data only when the data appear to be natural in the largest extent. We find that it is the case for fermion masses when the distribution is approximately Weibull. Such requirement can also give approximate values of $k$ and $l$. On the other hand, this process cannot be reversed. That is, we cannot use probability distributions to predict any precise mass values, since they are just randomly sampled from the distributions.

$\bullet$ There are also subtleties on choosing the data to be fitted. We assign the six quark masses into the same distribution because it is natural to expect that data from the same physics should come from the same distribution. However, again, the converse of this statement is not true. Here we separate quarks and charged leptons into different distributions to show a clearer physical picture, but it is mathematically consistent to ``design'' a distribution fitting both types of masses, although less excellently. Therefore data from the same distribution do not necessarily have the same physics.

$\bullet$ In this paper, we simply use the quark masses quoted in \cite{Tanabashi:2018oca}, which are the current quark masses instead of the constituent quark masses. We envision the quark masses in the absence of the strong QCD confining interaction (say, the pole masses as in the case of leptons).
However, since the QCD interactions are strong, especially for the light quarks, and quarks are confined, the determination of their masses is subject to the specific approach adopted, namely, lattice gauge theory or chiral perturbation theory (hence $\overline{MS}$ masses) for light quarks and pole masses for heavy quarks. Although the conversion between two types of masses is not fully understood, their differences are within a factor of two for the light quarks and within a few percent for the top quark. Fortunately, because of the very wide spread of the quark masses (from $u$ to $t$ quark), and the statistical nature of our analysis, such uncertainties in the quark mass determination has little impact in our conclusion.

$\bullet$ We are looking for a probability distribution which diverges at zero, is monotonic, normalizable and has no scale. The simplest such a distribution is the Weibull distribution \eqref{Weibullu}, where the divergence is power-like. For the probability distribution to be normalizable, the power $(k-1)$ must be weaker than $-1$, i.e., $k>0$. We can consider more non-trivial distributions like that obtained for $\Lambda$ \eqref{asymptotics of PDF}, where there are additional logarithmic factors. However, the logarithmic factors are weaker than the power factor, so they modify only slightly the divergent behavior at $u=0$. Considering the uncertainties in the quark masses and the sparseness of data, we can safely ignore these modifications to the Weibull distribution.

 \subsection{Neutrino Masses}

One would expect neutrino masses to follow the same kind of probability distribution but with a different $l$. Unfortunately, we do not have observed values for neutrino masses and cannot check this explicitly. Nonetheless, given our limited knowledge on neutrino masses, we will see that the proposal just suggested yields interesting insights on the values of neutrino masses. Namely, we show that the normal hierarchy is strongly preferred over the inverted hierarchy. Furthermore, we are able to obtain a very tight bound on the sum of the neutrino masses, $\sum m_{\nu} < 0.066$ eV, which implies that the lightest neutrino mass is of order of $10^{-3}$ eV or smaller.

Let us first recall the data. Take the three neutrino masses $m_\nu$ to be $(m_1,m_2,m_3)$ with $m_1<m_2<m_3$. The study of neutrino oscillations provides us with the square of the differences between two neutrino masses, $\Delta m^2_{ab}=m^2_a-m^2_b$. Depending on the choice of normal/inverted hierarchy, we have \cite{Capozzi:2017ipn}:
\begin{align}
 \label{normalHierarchy}
 \nonumber\text{Normal Hierarchy:} \quad 
 &\Delta m^2_{21}=(7.37^{+0.17}_{-0.16})\times10^{-5}\,{\rm eV}^2\,, \\
 &\Delta m^2_{31}=(2.56^{+0.04}_{-0.03})\times10^{-3}\,{\rm eV}^2 ; \\
 \label{invertedHierarchy}
  \nonumber\text{Inverted Hierarchy:} \quad 
  &\Delta m^2_{32}=(7.37^{+0.17}_{-0.16})\times10^{-5}\,{\rm eV}^2\,,\\
 &\Delta m^2_{31}=(2.54^{+0.04}_{-0.03})\times10^{-3}\,{\rm eV}^2.
\end{align}
The normal hierarchy is slightly preferred by cosmological data \cite{Xu:2016ddc} and further data from neutrino oscillations \cite{Simpson:2017qvj}. 
Following our proposal, since $P(m_\nu)$ peaks at $m_\nu=0$ and is monotonically decreasing with $m_\nu$, the normal hierarchy is clearly preferred over the inverted one. To be precise, when we randomly sample three neutrino masses from the distribution, the probability of getting a normal hierarchy (with similar but different $\Delta m_{ab}^2$) is much higher than that of the inverted hierarchy. Indeed, given the hierarchies above, we can extract the values of $r=m_{50}/\bar{m}=m_2/\bar{m}$ in both cases, as a function of $m_2$.
\begin{align}
\label{rN}
&r_{\rm norm}= \frac{3 m_2}{m_2+\sqrt{m_2^2-7.37\times10^{-5}}+\sqrt{m_2^2-7.37\times10^{-5}+2.56\times10^{-3}}} <1\,, \forall m_2 \\
&r_{\rm inv}= \frac{3 m_2}{m_2+\sqrt{m_2^2+7.37\times10^{-5}}+\sqrt{m_2^2+7.37\times10^{-5}-2.54\times10^{-3}}} >1\,, \forall m_2 \,,
\end{align} 
which are plotted in Figure \ref{rNrI}. 

\subsubsection{Dirac neutrinos}

The minimum value for \eqref{rN} is $r_{\rm norm}^{min}=0.435$, which corresponds to having a Weibull distribution with $k=0.668$ according to \eqref{peaking}. 
\begin{figure}[h]
 \centering
 \includegraphics[width=0.6\textwidth]{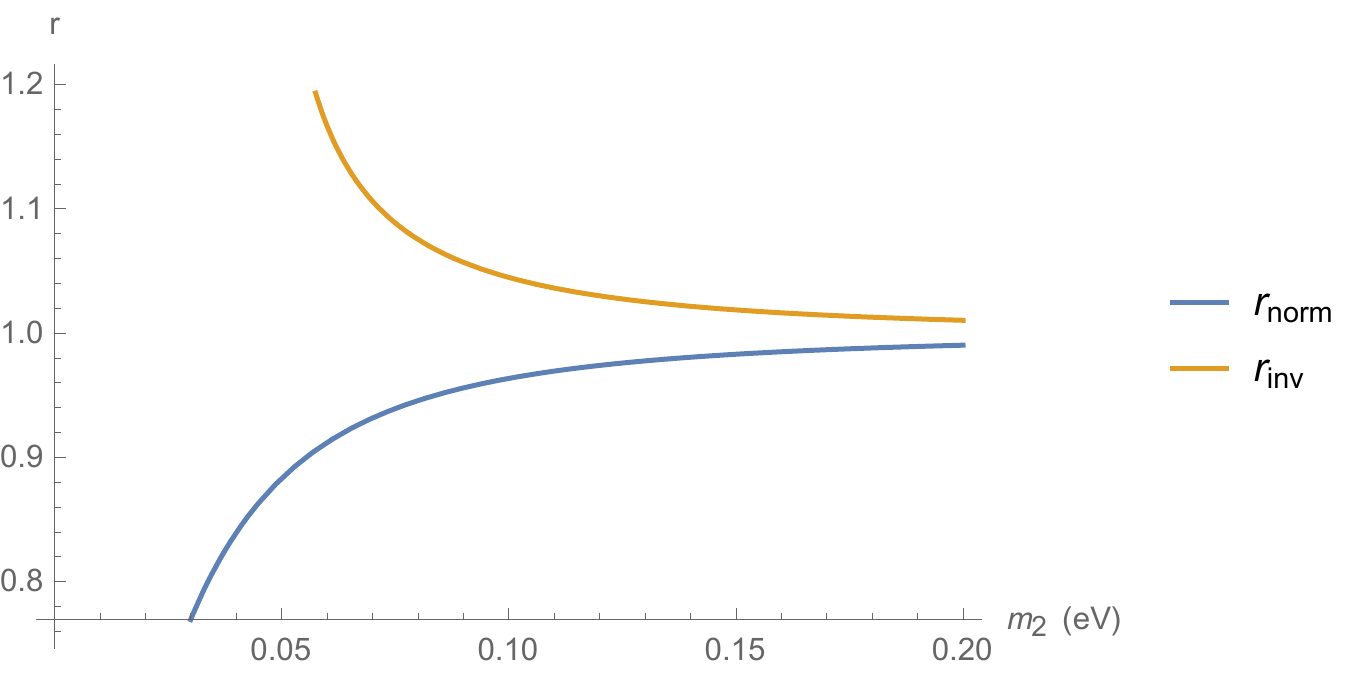}
 \caption{\label{rNrI} \small{The median/mean ratio $r$ as a function of median ($m_2$) for both normal and inverted hierarchies. The range of $m_2$ is bounded by the vanishing of $m_1$ (lower bound) and the observational bound on the sum of the neutrino masses $\sum m_{\nu} < 0.14$ eV  (upper bound). Our proposal on the probability distributions only allows $r<1$, so the inverted hierarchy is clearly ruled out. 
 }}
\end{figure}
Since the origin of the Dirac masses for the neutrinos should be similar to that for the the other fermions, we only allows $r<1$ so the inverted hierarchy is clearly ruled out, as shown in Fig.~\ref{rNrI}.

\begin{figure}[h]
 \centering
 \includegraphics[width=0.5\textwidth]{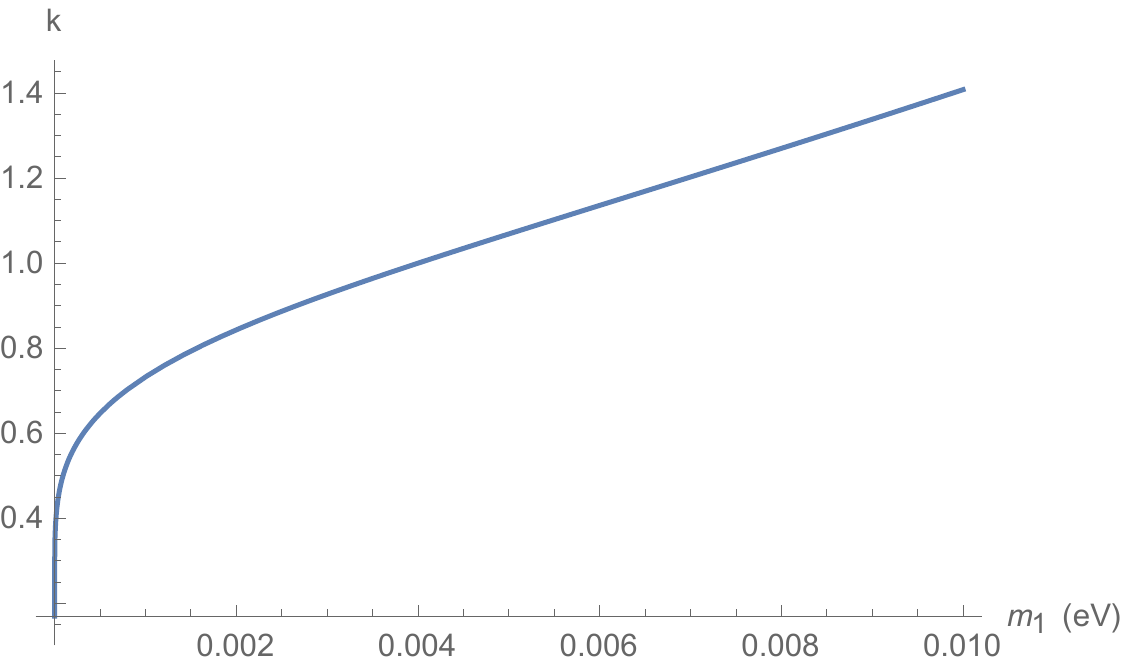}
 \caption{\label{bestfitk} \small{The best fit $k$ (in the Weibull distribution (\ref{Weibull})) as a function of the lightest neutrino mass $m_1$. To fit our proposal, we require $0< k <1$, hence $m_1<0.005 \, \text{eV}$.}}
\end{figure}

\begin{figure}[h]
 \centering
 \includegraphics[width=0.6\textwidth]{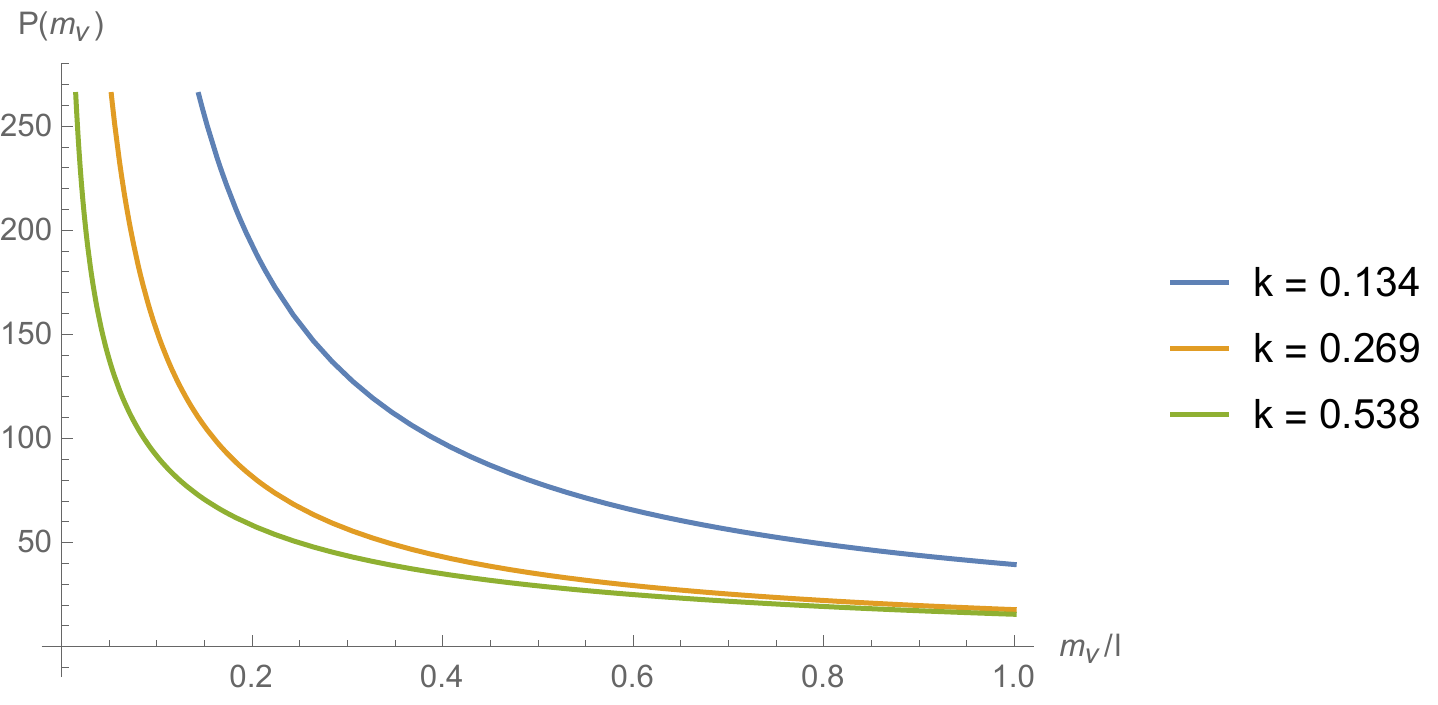}
 \caption{\label{kPlot} The plots of best fit $P(m_\nu)$ with $(k,l)=(0.134,0.0013\,\text{eV})$, $(0.269,0.0056\,\text{eV})$, $(0.538,0.013\,\text{eV})$.}
\end{figure}

Now that we have clearly seen that the inverted hierarchy is ruled out, we can ask how the normal hierarchy fits our proposal. Namely, we expect the best probability distribution fitting the three neutrino masses in the normal hierarchy should peak at zero but not elsewhere. Let us assume that the best fit is still a Weibull distribution, \eqref{Weibull}. 
As explained at the beginning of this section, to be consistent with our conjecture on $P(m_\nu)$, we need $k<1$.
Interestingly, this simple requirement imposes a tight upper bound on neutrino masses. This is because when the neutrino masses are too large, they become much closer one another than to zero, so that a probability distribution peaking at zero is not a good fit. Now, given one of the neutrino masses, let say $m_1$, we can determine all masses in the normal hierarchy, hence the best fit $P(m_\nu)=f(m_\nu;k,l)$. The resultant $k$ is then plotted in Fig.~\ref{bestfitk}. Now we see that requiring $k<1$, yields
\begin{equation}
 \label{nuB}
m_1<0.005 \,\text{eV},  \quad \quad \sum m_\nu< 0.066  \, \text{eV} \,.
\end{equation}
In principle, if neutrinos are Dirac fermions we should expect $k\simeq0.269$, while we should allow some uncertainty in $k$. As an example, if we impose the bound that $k$ should not be bigger than double the fitted value of $k=0.269$, i.e., $k <0.538$, we then find that
\begin{equation}
 \label{nuB2}
m_1<0.0002\,\text{eV},  \quad \quad \sum m_\nu < 0.061 \, \text{eV} \,.
\end{equation}
Comparing to the lowest value for the upper bound given by experiments, $\sum m_\nu<0.14\, \text{eV}$ 
(or a tighter bound but with less confidence, $\sum m_\nu<0.09\, \text{eV}$) \cite{Tanabashi:2018oca}, our result is clearly much stronger. 
If we demand that the same $k=0.269$ applies to the neutrino mass distribution $P(m_{\nu})$ as one expects for Dirac masses, we find that
\begin{equation}
 \label{num}
 m_1 \simeq 10^{-7}\, \text{eV}, \quad \quad  \sum m_{\nu} \simeq 0.0592^{+0.0005}_{-0.0004} \, \text{eV}.
 \end{equation}
 where $m_2 \simeq  0.0086$ eV and  $m_3 \simeq 0.051$ eV.
The resultant $P(m_\nu)$ with the above values of $k$ are plotted in Fig.~\ref{kPlot}. Note that the distributions $P(m/l) = f(u=m/l; k=0.269)$ are identical for the quarks, the charged leptons and the Dirac neutrinos. We emphasize that this result here strongly depends on the reliability of the probability distribution behavior proposal based on the string theory ``no parameter" property; so measuring the neutrino masses should provide a hint on how string theory can shed light on phenomenology.

\subsubsection{Seesaw Mechanism}

The above $P(m_{\nu})$ is for Dirac neutrino masses $m_D$ only. The smallness of the neutrino masses strongly suggests that a neutrino also has a large Majorana mass $M_M$. If the neutrino mass depends on a single flux parameter, $m_{\nu}(a)$, this results in a neutrino mass taking the form as $m_{\nu} \to 0$,
$$m_{\nu} (a) \simeq m_{D}(a)^2/M_M \simeq a^{1/k_S} = a^{2/k}$$
where for simplicity, we use Eq.(\ref{powerk})  to find the new shape parameter $k_S =k/2=0.134$ for the Seesaw case.  (Note that it is relatively insensitive to the flux dependence of $M_M$.) A smaller $k_{S}$ means a smaller ratio for the Seesaw, $r_S$, so the inverted hierarchy is more strongly ruled out.  Now, a best fit yields
\begin{equation}
P(m_{\nu}) \sim f(m_{\nu}; 0.134, 0.0013 \,\text{eV})
\label{nupM}
\end{equation}
where $l=0.0013$ eV.
Here we find that the lightest neutrino mass is now $m_1 \sim 10^{-8}$ eV, while the sum (\ref{num}) essentially remains unchanged. The corresponding $P(m_\nu)$ is plotted in Fig.~\ref{kPlot}.

Using the lepton mass scale $l_l=164$ MeV from the charged lepton distribution $P(m_l)$ (\ref{PofL}) and the neutrino mass scale $l_{\nu}=0.0013$ eV from the neutrino distribution $P(m_{\nu})$ (\ref{nupM}), we find that
\begin{equation}
\label{M_M}
M_M = l_l^2/l_{\nu} \simeq 2.1 \times 10^{16}\,  \text{GeV} \simeq M_S
 \end{equation}
 which is very close to the string scale obtained earlier \cite{Andriolo:2018dee}, which in turn is close to the GUT scale.

\section{Revisiting the Racetrack K\"ahler Uplift Model}
\label{Sec:Kuplift}

Let us revisit the Racetrack K\"ahler uplift model studied in Ref.~\cite{Sumitomo:2013vla,Andriolo:2018dee}. Here we like to discuss the robustness of the model in two aspects: \\
(1) So far we have focused on a single K\"ahler modulus case. Ref.~\cite{Sumitomo:2013vla} considers the multi-K\"ahler moduli case and check whether the multi-K\"ahler moduli case is compatible with the large volume approximation in
the Racetrack K\"ahler uplift and show that qualitatively the sharp peaking of $P(\Lambda)$ is maintained. Here we like to make  a few minor comments on the multi-K\"ahler moduli case. \\
(2) In going from the K\"ahler uplift model to the Racetrack K\"ahler uplift model, i.e., going from one non-perturbative term to two non-perturbative terms for the K\"ahler modulus in the superpotential $W$, we find that  $P(\Lambda)$ becomes substantially more peaked at $\Lambda=0$, resulting in a naturally small $\Lambda$. Here we check what happens if we go further, in including more non-perturbative terms in the single K\"ahler modulus case. Not surprisingly, we find that the peaking of $P(\Lambda)$ does not change qualitatively. 
 
We consider a 6-dimensional Calabi-Yau (CY) manifold $M$ with ($h^{1,1}$) K\"ahler moduli $T_j$ and $h^{2,1}>h^{1,1}$ number of complex structure moduli $U_i$, so the manifold $M$ has Euler number $\chi(M)=2(h^{1,1}-h^{2,1}) <0$.
This simplified model of interest is given by (setting $M_P=1$) \cite{Rummel:2011cd,deAlwis:2011dp,Sumitomo:2012vx,Sumitomo:2013vla},
\begin{equation}
 \begin{split}
  &V = e^{K} \left(K^{I \bar{J}} D_I W D_{\bar{J}} {\overline W} - 3\left|W \right|^2\right),\\
  &K = K_{\rm K} + K_S + K_U = -2 \ln \left({\cal V} + {\xi \over 2} \right) -   \ln \left(S+\bar{S} \right) -  \sum_{i=1}^{h^{2,1}} \ln \left(U_i + \bar{U}_i  \right),\\
  &{\cal V} \equiv {\vol \over \alpha'^3 } =  (T + \bar{T})^{3/2},  \quad 
  \xi =  -\frac{\zeta(3)}{4\sqrt{2}(2\pi)^3} \chi(M) \left( S + \bar{S} \right)^{3/2}>0, \\
  &W =  W_0(U_i,S) +  W_{NP}, \quad   W_{NP} = \sum_{i=1}^{N_g} A_i e^{-a_i T}
 \end{split}
 \label{LVS}
\end{equation}
Here, $M_P^2 \simeq {\cal V}/\alpha'$. The superpotential $W_0(U_i,S)$ for $U_i$ and the dilaton $S$ can in principle be extended to include other fields such as the Higgs boson. The uplift to deSitter space and the breaking of supersymmetry are provided by the $\alpha '$ correction $\xi$ term \cite{Becker:2002nn,Bonetti:2016dqh}.

\subsection{Multi-Non-Perturbative terms} 

The non-perturbative term $W_{NP}$ for the K\"ahler modulus $T$ in $V$ (\ref{LVS})  can be extended to the multi-K\"ahler moduli case. Ref.~\cite{Sumitomo:2013vla} considers the Swiss-Cheese case with two K\"ahler moduli in the Racetrack K\"ahler uplift model. It is shown that the sharp peaking of $P(\Lambda)$ at $\Lambda=0$ remains in the presence of additional K\"ahler modulus. It is clear from the analysis that, under reasonable circumstances,  more K\"ahler moduli will not change the overall picture of the single Racetrack K\"ahler uplift model. We may view this in another way. If the additional 
K\"ahler modulus is relatively heavy with respect to the other moduli and the dilaton, we may integrate it out in the low energy effective potential $V$. Its presence in $W_{NP}$ will be a function of the remaining light fields and some flux parameters, which may be approximated by the superpotential $W_0$ in $V$ (\ref{LVS}). Note that introducing higher $\alpha'$ corrections will change the value of $\xi$, but not the overall picture.

The non-perturbative term $W_{NP}$ for the K\"ahler modulus $T$ is introduced by gaugino condensates in the superpotential $W$ to stabilize the $T$ modulus \cite{Kachru:2003aw}. So far we have considered only the case with two non-perturbative terms ($n_{NP} = 2$) which form the Racetrack. However, in general there may be more gauge symmetries. To show the robustness of the model we introduce $n_{NP}$ non-perturbative terms, with the coefficients $a_i=2\pi/N_i$ for $SU(N_i)$ gauge symmetry ($i=1,...,n_{NP}$). The dependence of $A_i$ (also functions of some flux parameters) on $U_i, S$ is suppressed. They are treated as independent (real) random variables with smooth probability distributions that allow the zero values, while the dilation $S$ and the complex structure moduli $U_i$ are to be determined dynamically, yielding $\mathcal{W}_0$.

In the large volume region, the resulting potential may be approximated to, with $T=t + i \tau$,
\begin{equation}
 \begin{split}
  &V \simeq \left(-{a_1^3 A_1 \, \mathcal{W}_0\,  \over 2}\right) \lambda (x,y), \\
  & \lambda (x,y) =  - {e^{-x} \over x^2} \cos y - \sum_{i=2}^{n_{NP}}{\beta_i \over z_i} {e^{-\beta_i x} \over x^2} \cos (\beta_i y) + {\hat{C} \over x^{9/2}},\\
  &x = a_1 t, \quad y=a_1 \tau, \quad   z_i = A_1/A_i, \quad \beta_i = a_i/a_1=N_1/N_i >1, \quad {\hat{C}} = -{3 a_1^{3/2} \calw_0 \, \xi \over 32 \sqrt{2} A_1},
 \end{split}
 \label{approxpot}
\end{equation}
where we have chosen $N_1=\max \{N_1,N_2,...,N_{n_{NP}}\}$. Here we can already see that in the large volume region, the behaviour of $\lambda$ is dominated by the first term and the terms with smallest $\beta_i$. We expect that further adding more non-perturbative terms makes little changes. Following Ref.~\cite{Sumitomo:2013vla,Andriolo:2018dee}, the stability conditions at $y=0$ give us the cosmological constant:
\begin{equation}
 \label{lambdasol}
 \Lambda \simeq \frac{64\sqrt{2}a_1^{3/2}A_1^2}{243\xi}x^{5/2}\left(e^{-x}+\sum_{i=2}^{n_{NP}}\frac{\beta_i^2 e^{-\beta_i x}}{z_i}\right)^2\simeq \frac{3 \xi\calw_0^2}{4 (2t)^{9/2}}\,,
\end{equation}
where $x$ is determined as
\begin{equation}
\label{xsol}
e^{-x}+\sum_{i=2}^{n_{NP}}\frac{\beta_i^3 e^{-\beta_i x}}{z_i}\simeq 0\,.
\end{equation}
As explained in Ref.~\cite{Sumitomo:2013vla}, we can analyze the probability distribution of the cosmological constant, $P(\Lambda)$. 
After randomizing $A_i$, $N_i$ (with upper bound $N_{max}\sim\calo(100)$ given by F-theory \cite{Louis:2012nb}) and $\calw_0$, we collect all the classically stable solutions and find the probability distribution $P(\Lambda)$. Note that for $n_{NP}\geq3$, Eq. (\ref{xsol}) cannot be solved analytically, but has always an unique solution for the typical regime of parameters considered in \cite{Andriolo:2018dee}. To get the analytical behaviour of $P(\Lambda)$, we can numerically fit $P(\Lambda)$ to some well-known probability distribution functions. It turns out that for $n_{NP}\geq3$ and small $\Lambda$, the Weibull distribution works very well. Remarkably it is the same as the case of fermion masses. On the other hand, the case of $n_{NP}=2$ is given by \cite{Sumitomo:2013vla}
\begin{equation}
 P(\Lambda) \stackrel{\Lambda \rightarrow 0}{\sim} {243 \beta_2^{1/2} \over 16 (\beta_2-1)} {1 \over \Lambda^{\beta_2+1 \over 2 \beta_2}  (-\ln \Lambda)^{5/2}}.
  \label{asymptotics of PDF}
\end{equation}
So for $\beta_2 \gtrsim 1$, we see that the diverging behavior of the properly normalized $P(\Lambda)$ is very peaked as  $\Lambda \rightarrow 0$. Note that the $(-\ln \Lambda)$ part is sub-dominant.

Setting the median $\Lambda_{50}$ equal to the observed $\Lambda \sim 10^{-122}M_P^4$, and recalling that the superpotential has mass dimension 3, Eq.(\ref{lambdasol}) yields a new mass scale $\bf m$ \cite{Andriolo:2018dee},
\begin{align}
|\calw_0| \simeq 10^{-51} M_P^3 \quad \Rightarrow \quad {\bf m} = |\calw_0|^{1/3} \sim 10^{-17} M_P \sim 10^2 \,{\rm GeV} \,
\label{Key2}
\end{align}
where the string scale $M_S$ is around the GUT scale \cite{Andriolo:2018dee}, 
\begin{equation}
\label{M_S}
M_S = M_P/(2t)^{3/4} \simeq 1.3 \times 10^{16} \text{GeV}.
\end{equation}
which is close to the Majorana mass $M_M$ (\ref{M_M}).

Note that the solution we have is a meta-stable minimum. There is another solution it can decay to: for $t \to \infty$, the system decompactifies to 10-dimensional Minkowski spacetime. Fortunately, its lifetime is much longer than the age of our universe \cite{Tye:2016jzi}.
With $P(\Lambda)$ (\ref{asymptotics of PDF}), we have $P(\Lambda) \sim \Lambda^{-(\beta +1)/2 \beta}$, so Eq.(\ref{lambdasol}) suggests that, for $\calw_0 \to 0$, 
\begin{equation}
\label{Pcalw}
P(\calw_0) \sim \calw_0^{-1/\beta}.
\end{equation}
which sharply peaks at $\calw_0=0$.

We now compare the peaking behaviour with different $n_{NP}$. Below we show the peaking behaviour when $n_{NP}=2,3$ only, since further increasing $n_{NP}$ simply causes the similar but even smaller changes. Here, $\Lambda_{50}$ is the median. Following \cite{Andriolo:2018dee}, we find very simple approximation formulae for $\Lambda_{10}$ and $\Lambda_{50}$ in Table \ref{Tab1}.
\begin{table}[h!]
\begin{center}
$\begin{array}{ccc}
\toprule
n_{NP} & \Lambda_{10} & \Lambda_{50} \\
\midrule
2 & 10^{1.57-1.91 N_{max}} & 10^{-2.61-0.59 N_{max}} \\
3 & 10^{0.78-1.84 N_{max}} & 10^{-1.72-0.95 N_{max}} \\
\bottomrule
\end{array}$
\end{center}
\caption{\label{Tab1} \small{Approximate formulae for $\Lambda_{10}$ and $\Lambda_{50}$, where $\xi \simeq 10^{-3}$.}}
\end{table}

\begin{table}[h!]
\begin{center}
$\begin{array}{ccccc}
\toprule
N_{max} & \Lambda_{10}^{n_{NP}=2} & \Lambda_{50}^{n_{NP}=2} & \Lambda_{10}^{n_{NP}=3} & \Lambda_{50}^{n_{NP}=3} \\
\midrule
65 &  0.263 \times 10^{-122}  &  1.1 \times 10^{-41} & 1.51 \times 10^{-119} & 3.39 \times 10^{-64} \\
202 & 5.62 \times 10^{-385} &  1.62 \times 10^{-122} & 1.26 \times 10^{-371} & 2.4 \times 10^{-194} \\
67 & 3.98 \times 10^{-127} & 7.24 \times 10^{-43} & 0.316 \times 10^{-122} & 4.27 \times 10^{-66} \\
126 & 8.13 \times 10^{-240} & 1.12 \times 10^{-77} & 8.71 \times 10^{-232} & 3.80 \times 10^{-122} \\
\bottomrule
\end{array}$
\end{center}
\caption{\label{Tab2} \small{Estimates for $\Lambda_Y$ for $n_{NP}=2$ and $n_{NP}=3$ non-perturbative terms in $W$. Here $\xi \simeq 10^{-3}$. The $n_{NP}=2$ results are from Ref.
\cite{Sumitomo:2013vla,Andriolo:2018dee}.}}
\end{table}
In Table \ref{Tab2}, we present four cases, namely $\Lambda_{10}, \Lambda_{50}$ matching the observed $\Lambda \sim 10^{-122} M_P^4$ for $n_{NP}=2, 3$. 
Here, we see that $\Lambda_{50}$ for $n_{NP}=3$ matches the observed value for maximum $N_{max}=126$ (i.e., $SU(126))$. It turns out the resulting new scale is still ${\bf m} \sim 10^2$ GeV. For such a $N_{max}$,
$\Lambda_{50}$ for $n_{NP}=2$ is still orders of magnitude too big, when compared to the observed value. 
We see in Table \ref{Tab1} that $P(\Lambda)$ is more peaked at $\Lambda=0$ as $n_{NP}$ is smaller, but less suppressed at relatively higher percentiles. As a result, for the median $\Lambda_{50}$ to match observation can be achieved with smaller $N_{max}$ by increasing the number of non-perturbative terms in the Racetrack.

We have seen that converting the single non-perturbative term in $W_{NP}$ to the two-term Racetrack model makes a big difference in the form of $P(\Lambda)$, which becomes much more sharply peaked at $\Lambda=0$.
Ref.~\cite{Sumitomo:2013vla} shows that the introduction of additional K\"ahler moduli does not change the picture qualitatively. The same happens when we introduce more non-perturbative terms, in the single K\"ahler modulus case,  though quantitatively, the peaking of $P(\Lambda)$ becomes sharper. These show the robustness of the nice qualitative feature of the Racetrack K\"ahler uplift model, while it has the flexibility of quantitatively adjusting to fit observations.

\subsection{Flux basin and attractive basin}
\label{Sec:af}

\begin{figure}[t]
\includegraphics[width=\textwidth]{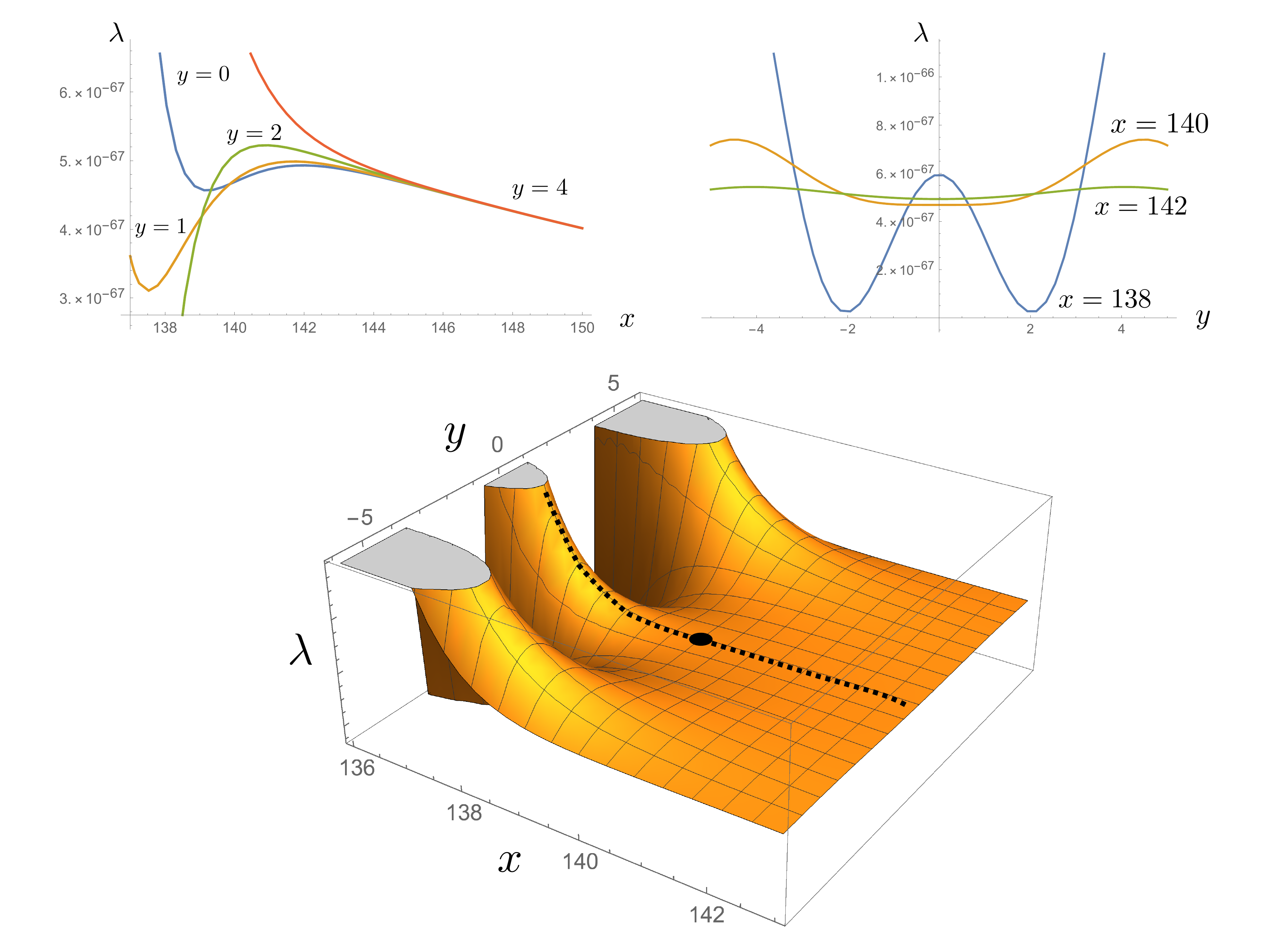}
\caption{\label{Fig:basinVfinal} \small{The plots for $\lambda(x,y=0,1,2,4)$, $\lambda(x=138,140,142,y)$ and $\lambda(x,y)$. The black spot represents the minimum.}}
\end{figure}
\begin{figure}[h]
\centering
\includegraphics[width=0.5\textwidth]{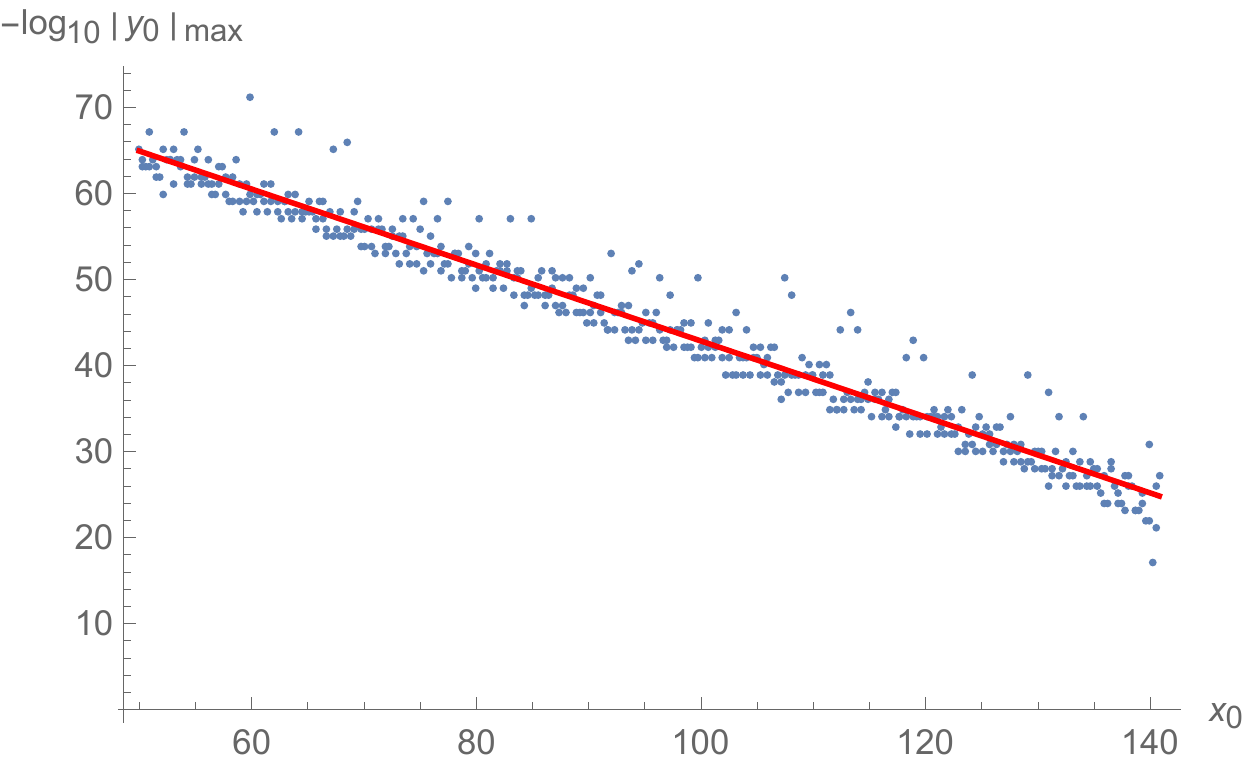}
\caption{\label{abr} \small{The attractive basin $\calb$ \eqref{attractivebasinV} for \eqref{parameters}}. In this example we use $H\sim10^{-14}$. For various $x_0$, we mark the ranges of $y_0$ by blue points, since there is large uncertainty by instability in numerical analysis. The red curve is a good linear fit to the results.}
\end{figure}
We now comment on the flux basin and attractive basin of our model in \cite{Andriolo:2018dee} ($n_{NP}=2$), to show how the universe can reach the vacuum we desire. As explained there, we see that the existence of a stable dS minimum with exponentially suppressed $\Lambda$ requires an exponentially suppressed $\calw_0$.
In the weak coupling regime, we would expect the moduli $U_i,S \simeq {\cal O}(1)$. Assuming the divergent behavior for $P(\calw_0)$ (\ref{Pcalw}), the discrete values of $\calw_0$ must be dense enough so that we can obtain a very small $\calw_0$ for $\Lambda$ to be small enough to fit observation. This may be achieved if there are enough number of flux parameters inside $W_0$ (a high dimensional flux basin), or the spacing of the few flux parameters in $\calw_0$ are small enough (dense discretuum in a low dimensional basin).

Let us now take a closer look at the attractive basin. For simplicity, since we are interested in the qualitative features, we will assume that the K\"ahler modulus $T$ rolls in the potential following the ``crude" equations of motion in an expanding universe:
\begin{equation}
 \label{crudeEOM}
\ddot{x}+3H\dot{x}+\frac{\partial V}{\partial x}=0 \,, \quad \ddot{y}+3H\dot{y}+\frac{\partial V}{\partial y}=0 \,,
\end{equation}
where $H$ is the Hubble parameter due to the expansion of the universe, and ``dot" means derivative with respect to an evolution parameter such as time. It is clear that the presence of $H$ can damp the rolling of $T$ and we assume a big enough $H$ to prevent over-shooting.  We further assume initially $\dot{x}=\dot{y}=0$ for simplicity, and denote the initial point be $(x_0,y_0)$. To proceed, we take the potential given by (see second column in Table \ref{Tab2}) 
\begin{align}
\nonumber&a \simeq 0.031\,, \quad 
A \simeq -0.1802\,, \quad 
\beta \simeq 1.005\,, \\
&z \simeq -0.504\,, \quad 
\hat C \simeq 2.49\times10^{-57} \,, \quad 
|\calw_0| \simeq 1.24\times10^{-51} \,,
\label{parameters}
\end{align}
which yields a minimum with $\Lambda \simeq 10^{-122}$ at $(x,y)\simeq(139.5,0)$, see Fig.~\ref{Fig:basinVfinal}.
Solving the differential equations \eqref{crudeEOM}, we can find the attractive basin, see Fig.~\ref{abr}
\begin{align}
\label{attractivebasinV}
\calb=(x_0,y_0) \quad\text{s.t.} \quad x_0 \lesssim142.5, \quad |y_0|_{max}=10^{-87.0+0.44x_0}\,.
\end{align}
As we see, in the $x$ direction, there is also a relative maximum at $x\simeq142.5$. If $x_0>142.5$, $T$ would roll to infinity, leading to decompactification and vanishing cosmological constant. So the de-compactified vacuum has a much bigger basin than that of the meta-stable vacuum. 

When $x<139.5$, the point $y=0$ is a relative maximum instead of minimum in $y$ direction. Therefore, if $|y_0|$ is too large, $T$ would roll away from $y=0$ and cannot reach the meta-stable minimum. It means that $T$ can roll backward in $x$ direction, leading to collapsing solutions. On the other hand, if $|y_0|$ is sufficiently small, $T$ rolls to $y=0$ when it passes the relative minimum in $x$ direction. When $x_0$ becomes larger, larger but still suppressed $|y_0|$ is allowed, since $T$ is closer to the relative minimum and escaping from $y=0$ becomes harder. Therefore $|y_0|_{max}$ increases as $x_0$ increases. Interestingly, the positive power $0.44$ still holds for other sets of parameters. Overall, there is an attractive basin in the neighborhood of the meta-stable minimum, as expected. 

\subsection{Relations to Swampland Conjectures}
\label{Sec:sc}

It has been recently pointed out that not every effective field theory one can write down can be embedded into string theory, and criteria have been conjectured to distinguish the string Landscape from the Swampland \cite{Vafa:2005ui}. Such considerations kicked off a review of all stringy inspired 4d EFT's. Here we would like to clarify where our model stands with respect to these claims.

\noindent\emph{No decoupled sectors in string theory:} As pointed out earlier, an EFT with two or more disconnected sectors cannot come from string theory, since the graviton and the dilaton will always couple them together. In the brane world scenario, the closed string modes will couple all sectors.  This point is crucial for a natural exponentially small $\Lambda$. In fact, suppose we allow for disconnected sectors, each one of these contributing $\Lambda_i$ to the total cosmological constant $\Lambda=\sum_i\Lambda_i$. Even if each sector yields a probability distribution $P_i(\Lambda_i)$ that peaks  (i.e. diverges) at $\Lambda_i=0$, the resulting peaking of $P(\Lambda)$ at $\Lambda=0$ is not granted, and it usually substantially weakened or erased. As we have shown, a sharp peaking at $\Lambda=0$ is necessary for the median $\Lambda$ value to match the observed value.

\noindent\emph{No dS conjecture:} Sometime ago, Dine and Seiberg proposed that there is no stable de Sitter vacuum in string theory in the \emph{asymptotic} regime of weak coupling $g_s\to0$ (or $s\to\infty$) \cite{Dine:1985he}. Our solutions, at weak but \emph{finite} coupling $g_s\sim\calo(10^{-1})$ ($s\sim\calo(1)$) are then untouched by this consideration. More recently, the observation of \cite{Dine:1985he} has been extended to any modulus (e.g., the volume modulus $t$), and it has been conjectured that string theory does not allow for (meta-)stable dS vacua at all \cite{Agrawal:2018own,Ooguri:2018wrx}. This is an extrapolation back from the asymptotic regime (e.g., $t\to\infty$), while explicit finite bounds are still absent. As before, our solutions at large but finite $t$ (and small but finite $g_s$) are consistent with the conjecture, and actually agree with it in the asymptotic $t\to\infty$ where we observe a runaway.

On the other hand, the corollary conjecture expressed in \cite{Dine:1985he,Agrawal:2018own} that eventual dS vacua will live in some \emph{strongly} coupled regime (emphasized also by \cite{Sethi:2017phn}) seem to clash with our findings. However, remember that non-perturbative interactive terms are crucial for the existence of de Sitter solutions at large finite volume $t$ and weak finite coupling $g_s$.  In particular, considering gaugino condensation, where $W_{NP}=\sum_i A_i e^{-2\pi T/N_i}+...$, we need the rank of the gauge group on condensing D7's to be $N_i\sim\calo(10^2-10^3)$ or bigger. Since usual non-perturbative terms are of the form $e^{-\phi/g}$ (for some field $\phi$ and coupling $g$), we see that in this sense the rank determines the strength of the coupling, so our model is strongly coupled, $g\sim N \gg 1$. Open questions are (a) whether non-perturbative terms in string theory really contribute to the EFT as the $W_{NP}$ above, and (b) whether such rank values are really achievable in string theory. While the first question has been recently addressed with a positive outcome in \cite{Carta:2019rhx,Bena:2019mte,Kallosh:2019oxv,Hamada:2018qef,Kachru:2019dvo}, the second one remains to be determined. 

\begin{figure}[t]
\centering
\begin{minipage}{\textwidth}
 \includegraphics[width=\textwidth]{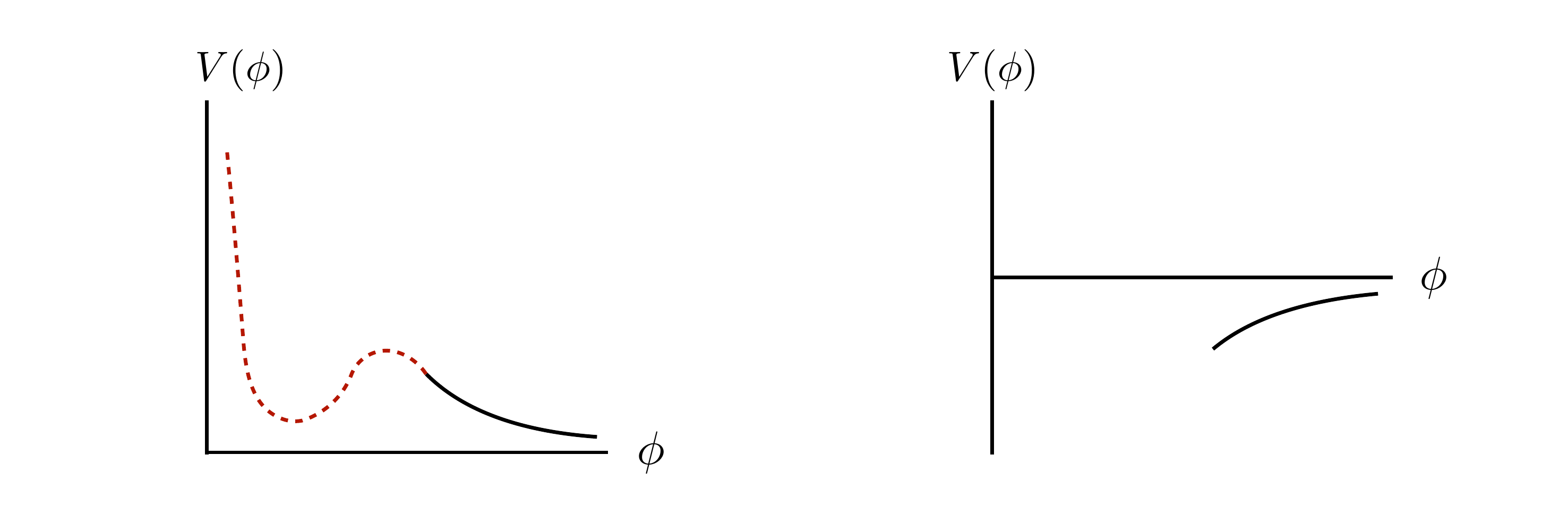}
 \caption{\label{Fig:DineSeiberg} \small{Dine--Seiberg proposals for the asymptotic behaviour of the effective potential in terms of a modulus $\phi$ (the dilaton in \cite{Dine:1985he}). The red dashed lines represent the possible dS vacuum, in what is expected to be a non-perturbative regime \cite{Dine:1985he}. Our model realizes this behaviour in the $\phi=t$ direction.}} 
 \end{minipage}
\end{figure}

\noindent\emph{Distance Conjecture:} Roughly speaking (see \cite{Brennan:2017rbf,Ooguri:2018wrx} for more detailed explanations), it states that when vevs of some moduli $\phi$ become large, towers of light particles with masses $m\sim e^{-\alpha \phi}$, for some constant $\alpha$, would emerge. In our model, complex structure and dilaton moduli are stabilised by fluxes, and they are expected to have vevs of the order $U_i,S\simeq {\cal O}(1)$. Therefore, these moduli are untouched by the distance conjecture. On the other hand, for our purposes, see Tab.~\ref{Tab2}, we need the K\"ahler modulus to be stabilised with $t\simeq {\cal O}(1000)$. As stated above, in absence of a clear bound, we do not consider this value to be part of the asymptotic regime, so there is no clash with any conjecture. However, if we consider this value as belonging to the asymptotics, we find agreement with the distance conjecture in that axion masses $m\sim e^{-at}$ exponentially decrease with $t$ (since $m\sim\Lambda^{1/2}$, we also find agreement with the conjecture suggested in \cite{Lust:2019zwm}). At the same time, however, we would violate the dS conjecture, and so the reasoning of \cite{Ooguri:2018wrx} should not apply or break down in our model.

\noindent\emph{Rolling to de Sitter:} We see in the above analysis that, for $y_0=0$, the only solution for $x_0>142.5$ (or $t=x_0/a=x_0N_{max}/2\pi >4582$ in $M_S$ units) is the runaway solution, where $t \to \infty$ as the universe decompactifies to a Minkowski space. For an exponentially small $\tau$ (i.e., $y_0$) and $x_0<142.5$, we are in an attractive basin so there are chances the universe will roll to the local de Sitter minimum at $(x_{min}, y_{min})=(139.5, 0.0)$, where the complex structure moduli as well as the dilaton take finite values. This happens for $y_0 < 10^{-87 +0.44x}$, otherwise the universe will roll away from the local minimum. 
Therefore, because of the very small attractive basin, a random sample of $(x_0, y_0)$ is very unlikely to hit a point that will roll to the de Sitter solution. If the universe starts with large values for the moduli (i.e., in the asymptotics), the``no de Sitter'' conjecture will reasonably say that it will never end in a de Sitter vacuum. Fortunately, the universe presumably starts with a large vacuum energy during the inflationary universe, so there is a non-trivial chance to pass through regions with finite modulus values. 

Notice the effective potential considered here is static, so the $V$ entering Eq.(\ref{crudeEOM}) may actually have different behavior for $(x_0, y_0)$ away from the minimum as the universe evolve towards the basin. It remains to be seen whether the change in $V$ will improve the chances of reaching the de Sitter vacuum.

\section{Discussions and Conclusion}
\label{Sec:discussion}

We believe the approach presented here, based on the probabilistic nature of string theory, can complement the ongoing discussion between Swampland and Landscape. Assuming that the dS Landscape is realised (with SM-like realisations belonging to it), we can identify some probabilistic features it should obey in order for our observed universe to be considered a \emph{statistically natural} realisation in it.  

Implementing the stringy constraints presented in Section \ref{Sec:toymodels}:(1) absence of free parameters; (2) all fields must couple together, directly or indirectly; and (3) a ``dense discretuum'' of flux values with smooth distributions that we scan over; we have proposed that the probability distribution $P(\calq)$ of (any low-energy flux-dependent quantity $\calq$ with nonnegative dimension) is smooth, and it is typically uniform or ``peaked at zero'', i.e., $\calq_{50}\ll\bar\calq$, while it cannot exhibit a monotonically increasing behaviour with $\calq$ (eventually peaking at some $\calq_{max}$).

\begin{figure}[h]
\centering
\begin{minipage}{\textwidth}
 \includegraphics[width=\textwidth]{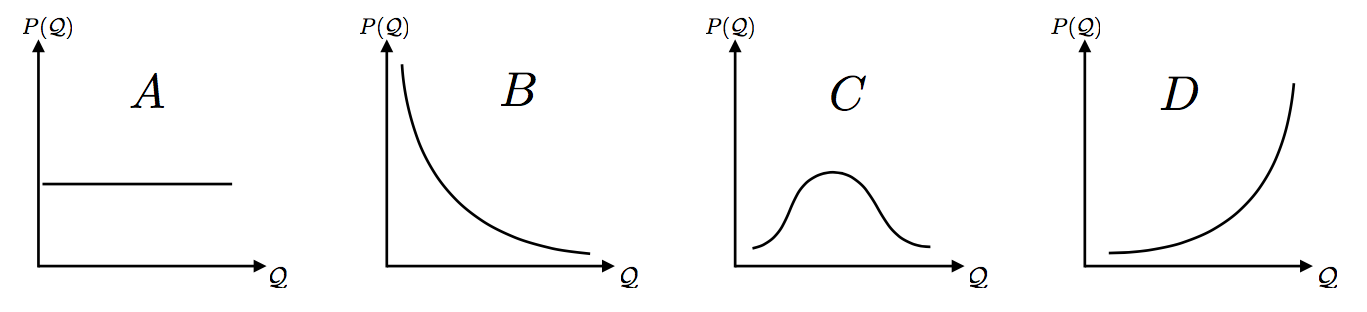}
 \caption{\label{Fig:pdfs} \small{Good and not so good distributions $P(\calq)$: type A,B are the most typical, C is allowed by some particular flux distribution or functional form that introduces a scale, while type D is not allowed by our proposal.}} 
 \end{minipage}
\end{figure}
Fig.~\ref{Fig:pdfs} illustrates a few possibilities. In $A$, the probability distribution $P(\calq)$ is flat, so there is no scale, or preference for any particular value. 
As we have shown in a number of cases (e.g., $P(\Lambda)$ (\ref{asymptotics of PDF}) and $P(m_f)$ (\ref{PofQ})), $P(\calq)$ peaks (diverges) at $\calq=0$, at times very sharply, as illustrated in $B$. 
That the peak always happens at $\calq=0$ is a simple probability theory property, and that string theory has no free parameter except the string scale (until new scales are generated dynamically) and so there is no preference for any specific scale when the quantity is well below the string scale. The degree of divergence of $P(\calq)$ at  $\calq =0$ is independent of any scale. Once we fix the dynamics (e.g., the value of $\Lambda$ and ${\bf m}$), scales can come into $P(\calq)$, but the degree of divergence will in general not change. This point is clearly illustrated by the quark and lepton mass spectra.

A  few comments are in order here : \\
$\bullet$ Consider the Planck mass $M_P$. Given the string scale $M_S$, we should  in principle obtain the probability distribution $P(M_P)$, given a specific K\"ahler uplift model. However, in the low energy effective theory approximation, we trust our analysis only when $M_P \gg M_S$, so $P(M_P)$ is a not particularly meaningful distribution to study in the present context. On the other hand, $P(1/M_P)$ may peak at  zero, i.e., $M_P \to \infty$, which implies decompactification. This is in agreement with the above discussion, that the attractive basin for decompactification is large compared to that for the meta-stable vacuum. Fortunately, tunneling from the meta-stable minimum to the decompactified solution takes much longer than the age of our universe \cite{Tye:2016jzi}.   \\
$\bullet$ A probability distribution of Type C can arise once scales are introduced dynamically, such as the value of $\Lambda$ and ${\bf m}$. It can also appear before that for a particular flux distribution. Here, the position of the peak is a new scale. It can appear as a function of the range of the discrete flux values we allow, e.g., the actual finite range, or the position of the peak and the width of a Gaussian distribution. The low energy effective theory is valid only for limited ranges of flux values. Going beyond will invalidate the particular low energy effective theory for a specific patch of the Landscape. Once a new scale appears in some flux distributions, distribution $P(\calq)$ of Type C  can also emerge. It is important that the measured fermion mass distributions agree with Type B but not Type C. \\
$\bullet$ For distribution $P(\calq)$ of Type B,  replacing a flat ($A$) flux distribution by a smooth ($C$) one does not impact on the degree of divergence of $P(\calq)$ at $\calq =0$, as long as the smooth distribution is non-vanishing at zero value and is nowhere divergent. The replacement of a flat flux distribution by a smooth distribution typically only affect the tail of $P(\calq)$, which changes the mean value $\bar \calq$ but has limited impact on the median $\calq_{50}$. \\
$\bullet$ If a quantity $\calq$ has negative dimension, the low-energy regime is at large $\calq$ instead of small $\calq$. In such regime, no interesting conclusions on the probability distribution $P(\calq)$ can be drawn, since we must have $P(\calq)\rightarrow 0$ when $\calq\rightarrow\infty$ for a normalizable $P(\calq)$. In particular, this is true even when $P(\frac{1}{\calq})$ diverges at $\frac{1}{\calq}\rightarrow 0$, which is required by our proposal. It can be understood as that the divergence of $P(\frac{1}{\calq})$ at one point is infinitely diluted in $P(\calq)$ at large $\calq$. On the other hand, since there is no reliable analysis in the high-energy regime, we cannot really determine the behaviour of $P(\calq)$ at small $\calq$.

Ref.~\cite{Sethi:2017phn} argues that any solution must have $\calw_0=0$ as the solution value of the superpotential $W_0$. In the Racetrack K\"ahler uplift model (\ref{lambdasol}), we find that this immediately implies that $\Lambda = 0$, i.e., the 10-dimensional Minkowski spacetime as the solution. Indeed, this de-compactification is the most natural solution, as a big enough value $t$ (which measures the compactified volume) will simply run away,  $t \to \infty$. As we have seen, the meta-stable minimum we find has a very small attractive basin compared to that for the de-compactified solution. So, intuitively, we may expect that it is highly unlikely to end in the meta-stable vacuum we are in today. However, if one believes in the inflationary universe scenario, our universe starts with a large vacuum energy density and rolls down. A very sharply peaked $P(\Lambda)$ implies that there is an exponentially large number of meta-stable solutions (with exponentially small $\Lambda>0$) for it to roll into, so ending in one of them is not as surprising as one's naive expectation suggests, since once it is trapped, it does not have a chance to explore the many more possibilities for it to decompactify. 

In the more realistic situation, we find that  $|\calw_0|=10^{-51} M_P^3$, which is a tiny shift away from zero, yielding an exponentially small positive $\Lambda$ that matches observation. Ref.~\cite{Lust:2005dy,Lust:2006zg} finds a $W_0(U_i, S)$ based on an analysis of orientifolds. In general, its value $\calw_0$ at the locally stable minimum solution can be very small, but non-vanishing. As explained in Section \ref{Sec:Kuplift}, we can also consider multi-K\"ahler moduli and a superpotential $W$ with only non-perturbative terms, i.e., $W = \sum A_j e^{-a_jT_j}$ only.  Now let us integrate out the heavy ones to reach the single K\"ahler modulus case (\ref{LVS}), so the non-perturbative terms of the heavy K\"ahler moduli inside $W$ are converted to terms that are function of $U_i$, $S$ and flux parameters, yielding a non-zero  perturbative looking $W_0(U_i, S, F^\alpha)$ inside $W$ (\ref{LVS}) (ignoring weak dependence on the light K\"ahler modulus). In short, a non-zero $W_0(U_i, S, F^\alpha)$ inside $W$ is generic, and such a term is generically non-zero when we sit at a minimum. 
 
We demonstrate that the application of string theory to particle physics phenomenology is possible even though we have no idea where the SM sits in the Landscape. We believe this novel statistical combination of top-down (the emergence of ${\bf m} \sim 10^2$ GeV) and the bottom-up (the fermion mass spectra) approaches provides a new way to do phenomenology. As our understanding of the Landscape improves, more phenomenology can be carried out, providing guidance in the search of the SM in the Landscape.
If our proposal is correct, it could easily explain why we observe a small cosmological constant, small couplings and the distribution of quark and lepton masses (as we have seen): we live in one of the \emph{statistically natural} vacua.

\section*{Acknowledgments}
We thank Yi Wang and Sav Sethi for several useful discussions. We also thank John Donoghue for informing us of their interesting works.

\bibliographystyle{utphys}
\bibliography{refs}

\end{document}